\documentclass[a4paper,11pt]{article}
\pdfoutput=1 

\usepackage{jheppub} 

\usepackage[makeroom]{cancel}
\usepackage[normalem]{ulem}

\title{A dark clue to seesaw and leptogenesis in {a pseudo-Dirac} singlet doublet scenario with (non)standard cosmology}

\author[a]{Partha Konar, }
\author[a]{Ananya Mukherjee, }
\author[a,1]{Abhijit Kumar Saha } 
\author{and}
\author[~a,b]{Sudipta Show}

\affiliation[a]{Physical Research Laboratory, Ahmedabad - 380009, Gujarat, India}
\affiliation[b]{Indian Institute of Technology, Gandhinagar - 382424, Gujarat, India}

\emailAdd{konar@prl.res.in}
\emailAdd{ananya@prl.res.in}
\emailAdd{aks@prl.res.in}
\emailAdd{sudipta@prl.res.in}

\abstract{We propose an appealing alternative scenario of leptogenesis assisted by dark sector which leads to the baryon asymmetry of the Universe satisfying all theoretical and experimental constraints. The dark  sector carries a non minimal set up of singlet doublet fermionic dark matter extended with copies of a real singlet scalar field. A small Majorana mass term for the singlet dark fermion, in addition to the typical Dirac term, provides the more favourable dark matter of pseudo-Dirac type, capable of escaping the direct search. Such a construction also offers a formidable scope to radiative generation of active neutrino masses. In the presence of a (non)standard thermal history of the Universe, we perform the detailed dark matter phenomenology adopting the suitable benchmark scenarios, consistent with direct detection and neutrino oscillations data. Besides, we have demonstrated that the singlet scalars can go through CP-violating out of equilibrium decay, producing an ample amount of lepton asymmetry. Such an asymmetry then gets converted into the observed baryon asymmetry of the Universe through the non-perturbative sphaleron processes owing to the presence of the alternative cosmological background considered here. Unconventional thermal history of the Universe can thus aspire to lend a critical role both in the context of dark matter as well as in realizing baryogenesis.\\

\vspace{.09cm}

}

\note{Note: During the review process of this manuscript, the affiliation of AKS has been changed to {\it``School of Physical Sciences, Indian Association for the Cultivation of Science, 2A and 2B Raja
S.C. Mullick Road, Kolkata 700 032"}. The new institutional email address of AKS is ``\textsf{psaks2484@iacs.res.in}".}

\begin{document}

\maketitle

\flushbottom

\section{Introduction}
Several cosmological challenges of particle physics keep us motivated to practice new proposals beyond the exiting ones. The first entity which wins our profound attention is the existence of dark matter~(DM) in the Universe. In spite of ample cosmological evidences, the origin and nature of dark matter still remain a mystery. And this mystery continues with the null results in several dark matter search experiments around the globe. It is conventional to state that the SM of particle physics lacks a viable candidate for dark matter. A plethora of beyond Standard Model (BSM) proposals have been cultivated already which are able to accommodate a stable dark matter candidate.  TeV scale LHC new physics search, together with celestial DM searches, keep rendering increasingly severe constraints on such models supporting cold DM in so-called Weakly Interacting massive particle (WIMP) paradigm~\cite{Arcadi:2017kky}. Therefore, the new challenge for theorists is to inquire after and trace the possible cause for the null results of DM at both direct search and collider experiments.

Similarly the understanding of the origin of the cosmological baryon asymmetry has been a challenge for both particle physics and cosmology. In an expanding universe,  baryon asymmetry can be generated dynamically by charge-conjugation (C),  charge-parity (CP) and baryon (B) number violating interactions among quarks and leptons. There are several attractive mechanisms which offer the explanation for the tiny excess of matter over antimatter, leptogenesis is one of a kind as pointed out by Fukugita and Yanagida \cite{Fukugita:1986hr} for the first time. In such a scenario the CP asymmetry is first generated in the lepton sector and later on gets converted into the baryon asymmetry via the non-perturbative sphaleron transition \cite{Rubakov:1996vz}. Leptogenesis via the out of equilibrium decay of the right handed neutrino (RHN) to SM leptons and Higgs in seesaw frameworks gained lots of attention in the last decade. For some earlier work one may look at refs. \cite{Pilaftsis:2003gt,Buchmuller:2004nz}. A prime aim of leptogenesis is that it can be used as a probe for the seesaw scale, thus opens up the testability of the heavy BSM particles responsible for generating tiny neutrino mass. Baryon asymmetry of the universe (BAU) is quantified as the ratio of the net baryon number density, $n_B$, to the photon density $n_\gamma$ and one can write \cite{ade2016planck}, 
\begin{equation}
 \eta_B^{\text{BBN}} = \frac{n_B - \bar n_B}{n_\gamma} = (2.6 - 6.2)\times 10^{-10}
\end{equation}

Since both dark matter and baryon asymmetry have cosmological origin, it is anticipated that there exists a correlation between the two. Indeed there have been a number of theoretical activities (see for instance \cite{Borah:2018rca,Yang:2018zrj,Biswas:2018sib} as some recent articles) which explore such an elegant connection. Majority of them have dealt with the standard thermal history of the Universe 
where it is assumed that the pre big bang nucleosynthesis (BBN) era was radiation dominated (RD). However, there is no direct evidence that obviate us from believing that prior to the radiation domination the Universe was populated by some other species. These non-standard scenarios must be consistent with the lower bound on the temperature of the last radiation epoch before BBN which is around $\mathcal{O}(1-10)$ MeV \cite{Kawasaki:2000en,Ichikawa:2005vw}. In modified cosmological scenario the expansion rate of the Universe naturally alters from what it is in case of the standard scenario. This could have considerable impact on standard description of particle physics phenomenology. Indeed, several exercises towards this direction have shown that in presence of such a non-standard history the DM phenomenology and the evolution of baryon asymmetry receive significantly deviation. For various model dependent and independent exercises on DM phenomenology in non-standard cosmology see \cite{Waldstein:2016blt,Redmond:2017tja,Binder:2017rgn,Dutta:2017fcn,DEramo:2017gpl,Hamdan:2017psw,Visinelli:2017qga,DEramo:2017ecx,Bernal:2018ins,Hardy:2018bph,Iminniyaz:2018das,Bernal:2018kcw,Arbey:2018uho,Biswas:2018iny,Betancur:2018xtj,Fernandez:2018tfa,Maldonado:2019qmp,Poulin:2019omz,Arias:2019uol,Allahverdi:2019jsc,Bernal:2019mhf,Chanda:2019xyl,Cosme:2020mck,Berger:2020maa,Han:2019vxi,Drees:2018dsj,Allahverdi:2018aux,PhysRevD.43.1063,Arbey:2008kv}. Some recent implications of non-standard  cosmology in the context of leptogenesis through RHN decay can be found in refs. \cite{Chen:2019etb,Mahanta:2019sfo,Abdallah:2012nm}. In the present framework we explicate the influence of such alternate cosmology in order to produce the observed BAU through the process of leptogenesis from the decay of a heavy SM singlet scalar.  

In this work our endeavor is to establish \emph{a comprehensive connection between dark sector and observed baryon asymmetry of the Universe in a non-standard cosmological scenario}. { The dark sector involves an extended version of the singlet doublet Dirac dark matter \cite{Yaguna:2015mva} framework} with the dark matter weakly interacting with the thermal bath. It was earlier shown by us \cite{Konar:2020wvl} that presence of a small Majorana mass for the singlet fermion in addition to the Dirac mass makes the DM (admixture of singlet and doublet) of pseudo Dirac nature \footnote{ In view of the rich phenomenology associated with a pseudo-Dirac DM, we deform the pure Dirac version of the singlet doublet DM model. One can find the Majorana version of the singlet doublet dark matter in \cite{Cohen:2011ec}. For other related works and associated phenomenology based on a similar kind of set up one can refer to \cite{Fiaschi:2018rky,Restrepo:2015ura,Bhattacharya:2017sml,Bhattacharya:2015qpa,Bhattacharya:2018fus,Barman:2019tuo,Arcadi:2018pfo,Calibbi:2018fqf,Esch:2018ccs,Maru:2017pwl,Maru:2017otg,Xiang:2017yfs,Abe:2017glm,Banerjee:2016hsk,Horiuchi:2016tqw,Calibbi:2015nha,Cheung:2013dua,Enberg:2007rp,DEramo:2007anh,Barman:2019aku,Restrepo:2019soi,Freitas:2015hsa,Cynolter:2015sua,Bhattacharya:2016lts,Bhattacharya:2016rqj,Wang:2018lhk,Abe:2019wku,Barman:2019oda}.}.{ The pseudo-Dirac dark matter is known to leave imprints at the collider in the form of a displaced vertex which can be traced. The pseudo Dirac nature also assists the DM to escape from the direct search experiments by preventing its interaction with the neutral current at the tree level \cite{DeSimone:2010tf}. We have shown that eventually the absence of a neutral current at the tree level leads to a substantial improvement for the allowed range of the mixing angle between the singlet and doublet fermion which was otherwise strongly constrained.} In \cite{Konar:2020wvl} we also extend the minimal singlet dark matter set up by inclusion of copies of a dark singlet scalar field to yield light active neutrino masses radiatively. We particularly have emphasized that the  Majorana mass term which is related to non observation of DM at direct search experiments can yield the correct order of light neutrino masses.
In the present work we explore the DM phenomenology in an identical set up by making an important assumption of presence of a non-standard thermal history of the Universe. In particular we consider the presence of a popular non-standard scenario before the BBN dubbed as fast expanding Universe \cite{DEramo:2017gpl}. 

As previously mentioned we also offer a slightly different approach for realizing leptogenesis, where the lepton asymmetry originates from the lepton number and CP violating decay of singlet dark scalar fields into SM leptons and one of the dark sector fermion.  The produced lepton asymmetry further can account for the observed baryon asymmetry of the Universe through the usual sphaleron process. We specifically have shown that the presence of a non-standard era in the form of a fast expanding Universe is slightly preferred in order to generate the observed amount of matter-antimatter asymmetry in this particular set up.

This work is organised as follows. In Section~\ref{model} we present the structure and contents of the model, which is primarily an extended version of the singlet doublet model. Theoretical as well as experimental constraints of the model parameters are debated in Section~\ref{constraints}.
Section~\ref{fastexpanding} is kept for explaining the cosmology of fast expanding universe where working mathematical forms are provided to utilise them in following sections. We detail the DM phenomenology in presence of non-standard cosmology in the Section~\ref{DM}. Different aspects of parameter dependance and related constraints are discussed quantifying the effect of non-standard scenario. In Section \ref{sec:neu}, we present the neutrino mass generation technique. Then Section~\ref{leptogenesis} is dedicated for the baryogenesis through leptogenesis and the required analytical formula realizing the same. Results and analysis for neutrino mass and BAU are shown in Section~\ref{resultsL}. Finally we summarize our findings and conclude in Section~\ref{conclusion}.

\section{Structure of the model} \label{model}
We propose a pseudo-Dirac singlet doublet \emph{fermionic dark matter} model 
and extend it minimally to accommodate \emph{neutrino mass} and \emph{baryon asymmetry of the Universe}. { The fermion sector in the set up includes one vector fermion singlet ($\chi=\chi_L+\chi_R$) and another $SU(2)_L$ vector fermion doublet ($\Psi=\Psi_L+\Psi_R$)}. The BSM scalar sector is enriched by three copies of a real scalar singlet field ($\phi_{1,2,3}$). We consider the SM fields to transform trivially under a imposed $\mathcal{Z}_2$ symmetry while all the BSM fields are assigned odd $\mathcal{Z}_2$ charges~(see Table~\ref{tab:particle}). The BSM fields are non-leptonic in nature.
\begin{table}[h!]
	\centering
	\begin{tabular}{ |p{5cm}|p{6cm}|p{.8cm}||p{.6cm}| }
		\hline
		\hspace{0.8cm}BSM and SM Fields& $\hspace{0.5cm}SU(3)_C\times SU(2)_L\times U(1)_Y\equiv \mathcal{G}$ &$\hspace{-0.1cm}{ U(1)_L}$&$\mathcal{Z}_2$ \\
		\hline
		\hline
		$\hspace{1.8cm}{ \Psi_{L,R}}$ 
	  & \hspace{0.9cm}1
		\hspace{1.4cm}2 \hspace{1.3cm}-$\frac{1}{2}$\hspace{1.1cm}&$\hspace{0.3cm}{ 0}$ & $-$ 
		\\
		\hline
		$\hspace{1.8cm}$ ${ \chi_{L,R}}$ $ $  & {\hspace{0.9cm}1 \hspace{1.4cm}1\hspace{1.57cm}0}\hspace{1.07cm} &$\hspace{0.3cm}{ 0}$& $-$ \\
		\hline
		$\hspace{1cm}{\phi_i~(i=1,2,3)} $  & {\hspace{0.9cm}1 \hspace{1.4cm}1\hspace{1.57cm}0}\hspace{1cm}& $\hspace{0.3cm}{ 0}$&$-$ \\
		\hline
		\hline
		$\hspace{0.9cm}\ell_L \equiv \begin{pmatrix}
		\nu_\ell  \\
		\ell
		\end{pmatrix} $  & \hspace{0.9cm}1
		\hspace{1.4cm}2 \hspace{1.3cm}-$\frac{1}{2}$\hspace{1.1cm} &$\hspace{0.3cm}{ 1}$& $+$ 
		\\
		\hline
		$\hspace{0.7cm}H \equiv \begin{pmatrix}
		w^+  \\
		\frac{1}{\sqrt{2}}(v+h+iz)
		\end{pmatrix} $  & \hspace{0.9cm}1
		\hspace{1.4cm}2\hspace{1.4cm}~$\frac{1}{2}$\hspace{0.97cm}&$\hspace{0.3cm}{ 0}$ & + \\
		\hline
	\end{tabular}
	\caption{Fields and their quantum numbers under the SM gauge symmetry, { lepton number} and additional $\mathcal{Z}_2$.}
	\label{tab:particle}
\end{table}
The Lagrangian of the scalar sector is given by
\begin{align}
\mathcal{L}_{\rm scalar}=|D^\mu H|^2+\frac{1}{2}(\partial_\mu\phi)^2-V(H,\phi),
\end{align}
where,
\begin{align}
D^\mu= \partial^\mu-ig\frac{\sigma^a}{2}W^{a\mu}-ig^\prime Y B^\mu,
\end{align}
with $g$ and $g^\prime$ stand for the $SU(2)_L$ and the $U(1)_Y$ gauge couplings respectively.
Below we write the general form of the scalar sector potential $V(H,\phi)$ consistent with the charge assignment in Table \ref{tab:particle}:
\begin{align}
V(H,\phi_i)=-{\mu_H^2}\, (H^\dagger H)+\lambda_H \, (H^\dagger H)^2+\frac{\mu_{ij}^2}{2} \, \phi_i\phi_j+\frac{\lambda_{ijk}}{2} \, \phi_i^2\phi_j\phi_k+\frac{\lambda_{i j}}{2} \, \phi_i \phi_j (H^\dagger H).
\end{align}
After minimization of the scalar potential in the limit $\mu_H^2,\mu_{ij}^2>0$ the vacuum expectation values (vev) for both the scalars $H$ and $\phi_i$'s can be obtained as given below,
\begin{align}
\langle H\rangle=v,~~\langle\phi_i\rangle=0.
\end{align}
For simplification, we consider $\lambda_{ij}, \lambda_{ijk}$ as diagonal in addition to mass matrix for the scalars, parameterized as Diag($M_{\phi_1}^2,M_{\phi_2}^2,M_{\phi_3}^2$). Since $\langle\phi_i\rangle=0$, $\mathcal{Z}_2$ remains unbroken which stabilizes the DM candidate.

The Lagrangian for the fermionic sector at tree level is written as:
\begin{align}
\mathcal{L}=\mathcal{L}_{f}+\mathcal{L}_{Y},
\end{align}
where,
\begin{align}
\mathcal{L}_{f}=i\bar\Psi \gamma_\mu D^\mu \Psi+i\bar\chi  \gamma_\mu\partial^\mu \chi-M_\Psi \bar\Psi \Psi-M_\chi \bar\chi \chi-\frac{m_{\chi_L}}{2}\overline{\chi^c}P_L\chi-h.c.-\frac{m_{\chi_R}}{2}\overline{\chi^c}P_R\chi-h.c.,\label{eqn:FeynA}
\end{align}
and
{\color{black}
\begin{align}
\label{eqn:Fyukawa}
\mathcal{L}_{Y}=Y_1\bar\Psi_L \tilde{H} \chi_R+Y_2\bar\Psi_R \tilde{H} \chi_L+h_{\alpha i} \bar\ell_{L_\alpha} \Psi_R \phi_i+h.c..
\end{align}
}
In the $\mathcal{L}_f$, the doublet has a Dirac like mass term $M_\Psi\bar{\Psi}\Psi$ which can be expanded as $M_\Psi(\overline{\Psi_L}\Psi_R+\overline{\Psi_R}\Psi_L)$. While for $\chi$ field both the Dirac $M_\chi (\overline{\chi_L}\chi_R+ \overline{\chi_R}\chi_L)$ and Majorana type masses ($m_{\chi_{L,R}}$) appear in Eq.~(\ref{eqn:FeynA}),  which is perfectly allowed by the imposed $Z_2$ symmetry. {\color{black}In a similar line the Eq.~(\ref{eqn:Fyukawa}) shows the Yukawa like
interaction pattern of  $\psi_{L,R}$ and $\chi_{L,R}$ with the SM Higgs and $\phi$. Hereafter we work with a generic choice $Y_1=Y_2\equiv Y$ in order to reduce the number of free parameters in the model (see \cite{Herrero-Garcia:2018koq,Cynolter:2008ea} for such an example). This particular choice of equality helps us to evade the spin dependent direct detection bound (please refer to footnote \ref{ft:axial}). With this equality the first two Yukawa terms can be written in a compact form like $Y \bar{\Psi}\tilde{H}\chi$.} 
We specifically assume that the Majorana mass for $\chi$  field is much smaller than the Dirac one \textit{i.e. $m_{\chi_{L,R}}\ll M_\chi$}. In the present framework the lightest neutral fermion is a viable dark matter candidate which is of pseudo-Dirac nature in the limit $m_{\chi_{L,R}}\ll M_\chi$. As we see in \cite{Konar:2020wvl} that this non-vanishing $m_{\chi_{L,R}}$ assists in evading strong spin-independent dark matter direct detection bound. In addition, it is also found \cite{Konar:2020wvl} to be crucial in generating light neutrino mass radiatively.

The presence of a non-vanishing $m_{\chi_{L,R}}$ and and $M_\chi$ along with $\phi$ being a real scalar field and non-vanishing coupling coefficient $Y$ result in symbolizing the Yukawa like interaction ($h$) involving SM leptons and the doublet $\psi$ as a lepton number violating vertex at tree level. The interaction of DM with the SM particles mediated through the Higgs is realized by the first term in Eq.~(\ref{eqn:Fyukawa}), whereas the second term which is also responsible for active neutrino mass generation through radiative loop \cite{Konar:2020wvl} manifests the explicit violation of the lepton number~\footnote{The purpose of choosing the dark sector scalar fields as real is justified to pave the way for explicit lepton number violation \cite{Restrepo:2015ura} in Eq.(\ref{eqn:Fyukawa}).}. 

\begin{figure}[t]
\centering
\includegraphics[height=7cm,width=9cm]{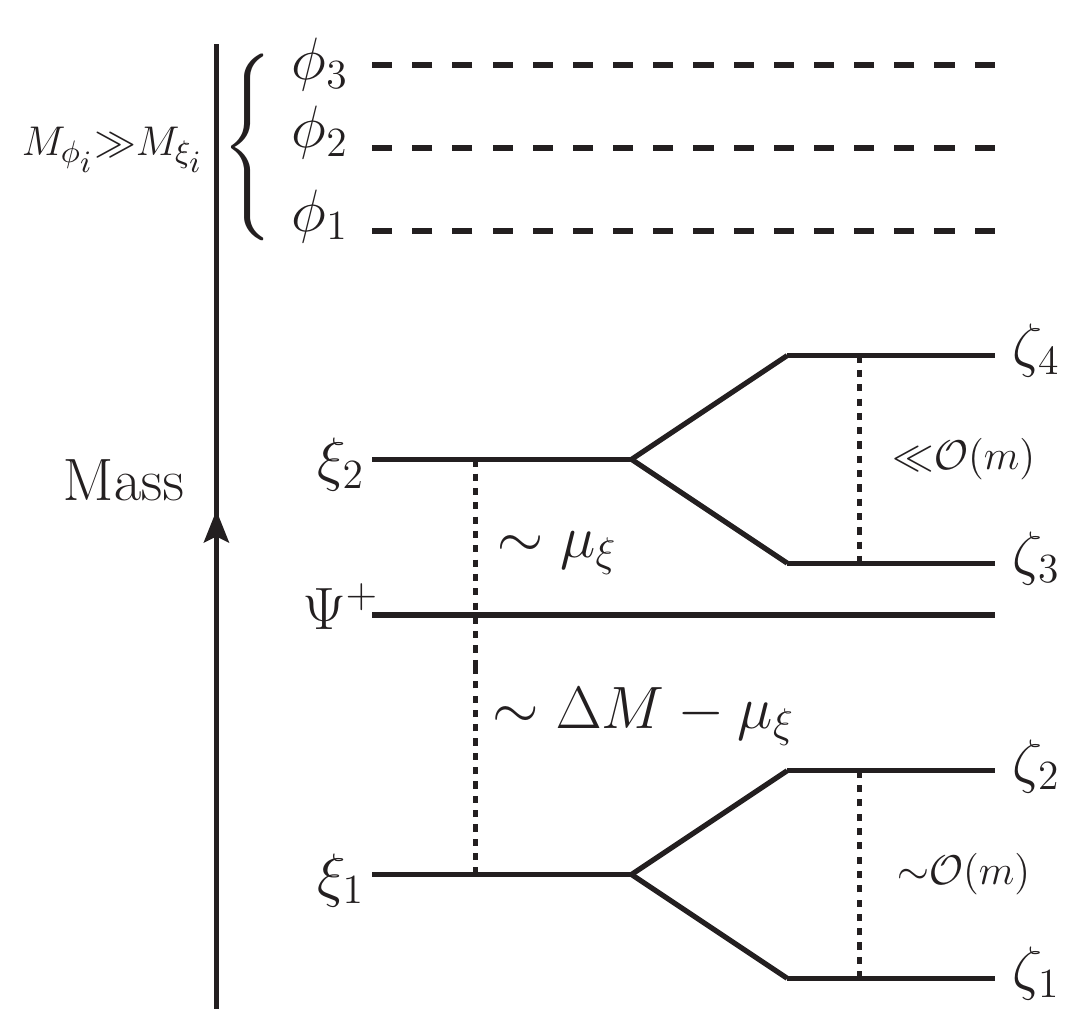}
\caption{Mass spectrum of the dark sector, showing the lightest pseudo-Dirac mode as the dark matter and other heavy BSM fermions and scalars. {\color{black} The mass of the charged fermion is $M_\Psi$ and it lies somewhere in between $\xi_1$ and $\xi_2$ with $\mu_\xi=\frac{\Delta M\sin^2(2\theta)}{4}$ in the limit of small $\sin\theta$. The mass ordering is subject to change depending on the numerical values of $m$, $\Delta M$ and $\sin\theta$.}} 
\label{massorder}
\end{figure}

 In the present study, we consider $M_{\phi_i}\gg M_\psi,m_{\chi_{L,R}}$ such that the role of $\phi$ fields in DM phenomenology is minimal \footnote{Ideally the scalars, being a part of the dark sector can engage in DM phenomenology through coannhilation processes however considering the mass pattern in Fig. \ref{massorder} their contributions turn out to be minimal.}. After the spontaneous EW symmetry breaking, the Dirac \footnote{  The Majorana version of the singlet doublet dark matter accommodates one pair of Weyl $SU(2)_L$ doublet fermions and one Weyl singlet fermion. Thus the number of neutral Weyl degrees of freedom is three. While in our case there exist four neutral Weyl degrees of freedom.} mass matrix for the neutral DM fermions is given by (in $m_{\chi_{L,R}}\rightarrow 0$ limit),
\begin{align}
 \mathcal{M}_D=\begin{pmatrix}
		M_\Psi & M_D  \\
		M_D & M_\chi 
		\end{pmatrix},
\end{align}
where we define $M_D=\frac{Yv}{\sqrt{2}}$. After diagonalisation of $\mathcal{M}_D$ the mass eigenvalues are computed as,
\begin{align}\label{massE}
&M_{\xi_1} = \frac{M_\chi+M_\Psi}{2} -\frac{1}{2} \sqrt{4 M_D^2+M_\chi^2-2 M_\chi M_\Psi+M_\Psi^2},\\
&M_{\xi_2} =\frac{M_\chi+M_\Psi}{2} +\frac{1}{2} \sqrt{4 M_D^2+M_\chi^2-2 M_\chi M_\Psi+M_\Psi^2},
\end{align} 
where the Dirac mass eigenstates are represented as $(\xi_1,\xi_2)$. It is evident from Eq.(\ref{massE}) that $\xi_1$ is the lightest eigenstate. The mixing between two flavor states, {\it i.e.} neutral part of the doublet ($\psi^0$) and the singlet field ($\chi$) is parameterised by $\theta$ as
\begin{align}
\sin 2\theta= \frac{\sqrt{2}~Yv}{\Delta M},
\end{align}
where $\Delta M =M_{\xi_2}-M_{\xi_1}$ which turns out to be of the similar order of $M_{\Psi}-M_{\chi}$ in the small $\theta$ limit.
Also, in small mixing case, $\xi_1$ can be identified with the singlet $\chi$. 
 In the limit $m\rightarrow 0$ where we define 
\begin{equation}\label{majorana}
m=(m_{\chi_L}+m_{\chi_R})/2,
\end{equation}
 the Majorana eigenstates of $\xi_1$ ({\it i.e.} $\zeta_1,~\zeta_2$) are degenerate. A small amount of non-zero $m_{\chi_{L,R}}$ breaks this degeneracy , and we can still write 
 \begin{align}
&\zeta_1\simeq \frac{i}{\sqrt{2}}(\xi_1-\xi_1^c),\label{eq:eigenS}\\
&\zeta_2\simeq \frac{1}{\sqrt{2}}(\xi_1+\xi_1^c).
\end{align}
in the pseudo-Dirac limit $m\ll M_{\zeta_1},M_{\zeta_2}$ where $M_{\zeta_1,\zeta_2}\simeq M_{\xi_1}\mp m$. In a similar fashion, the state $\xi_2$ would be splitted into $\zeta_3$ and $\zeta_4$. Hence we will have four neutral  mass eigenstates in the DM sector with the lightest state ($\zeta_1$) being the DM candidate. { Since all of the mass eigenstates have pseudo-Dirac origin, we mark them as ``pseudo-Dirac" states. For a formal understanding on the construction of pseudo Dirac fermion in terms of the Weyl spinors, we refer the readers to Appendix \ref{pseudo}} .

For a representative mass spectrum of the dark sector, please follow Fig. \ref{massorder}, showing the lightest pseudo-Dirac mode as the dark matter candidate together with other heavy BSM fermions and scalars. In the following section we look into the possible constraints before emphasizing cosmological predictions of the model.
\section{Model Constraints}\label{constraints}
In this section we summarize the possible constraints on the model parameters arising from different theoretical and experimental bounds.

\begin{itemize}
\item \textbf{Perturbativity and stability bounds:} Any new theory is expected to obey the perturbativity limit which imposes strong upper bounds
on the model parameters:
\begin{align}
\lambda_{ij},~\lambda_{ijk}< 4\pi,~~{\rm and}~~ Y,~h_{ij}<\sqrt{4\pi}.
\end{align}
It is also essential to ensure the stability of the scalar potential in any field direction. The stable vacuum of a scalar potential in various field directions are determined by the co-positivity conditions \cite{Chakrabortty:2013mha,Kannike:2012pe} where all the scalar quartic couplings are involved. Here we are considering all the scalar quartic couplings as real and positive and thus automatically satisfy the necessary co-positivity conditions.
 
\item \textbf{Bound on Majorana mass parameter:} In the presence of a small Majorana mass, the $\xi_1$ state gets splitted into two 
non degenerate Majorana eigenstates. This triggers the possibility of inelastic scattering of $\xi_1$ with nucleon to produce $\xi_2$.
Such inelastic scattering would give rise to non zero excess of nucleon recoil into direct detection experiments ({\it e.g. }XENON 1T) which is strongly disfavored. Hence, it is recommended to forbid such kind of inelastic processes. This poses some upper limit on the Majorana mass parameter
$m_{\chi_L}+m_{\chi_R}\gtrsim 240$ KeV for DM having mass $\mathcal{O}(1)$ TeV considering Xenon detector \cite{TuckerSmith:2001hy,Hall:1997ah}.

\begin{figure}[t]
\centering
\includegraphics[height=5cm,width=7cm]{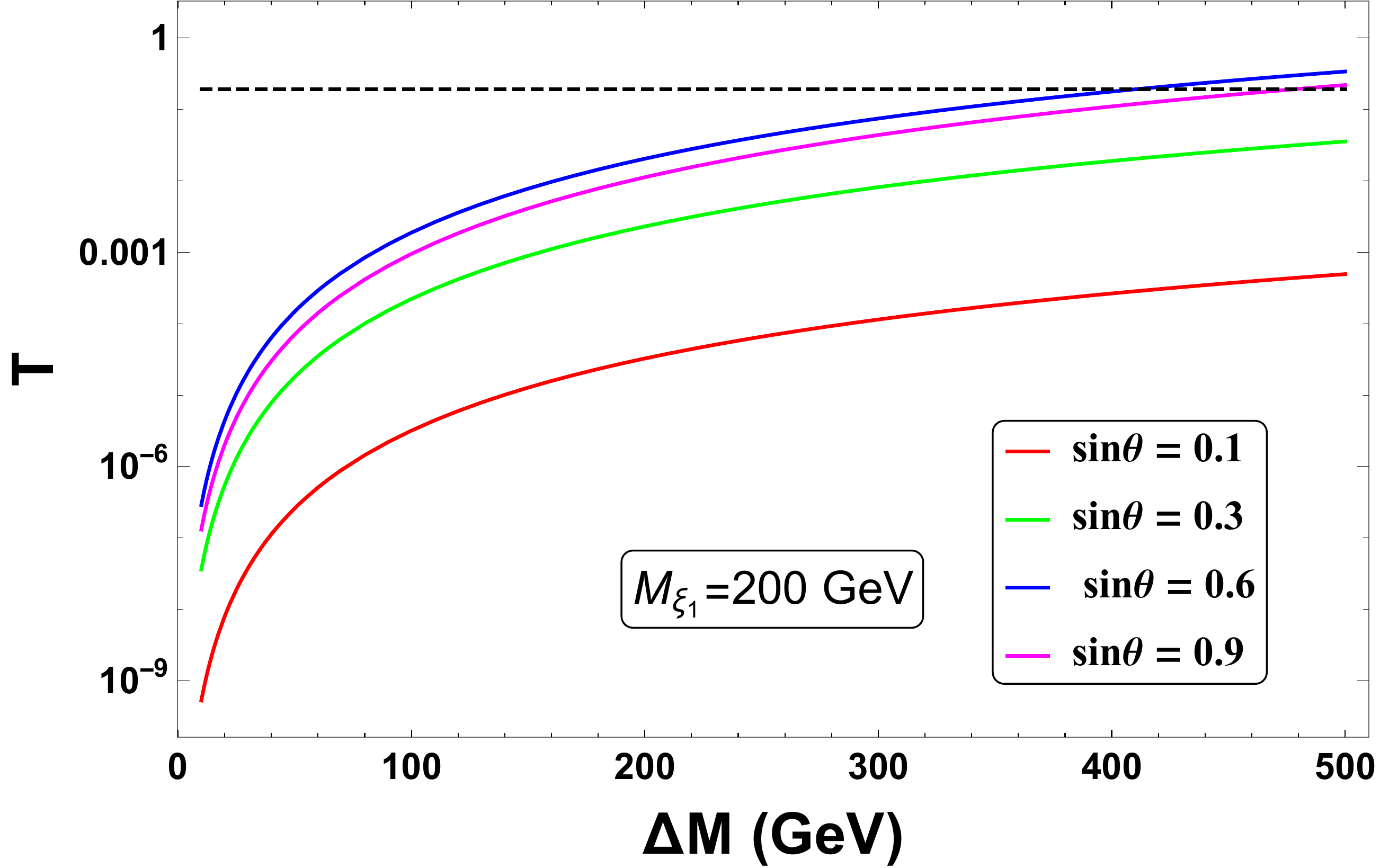}~~~
\includegraphics[height=5cm,width=7cm]{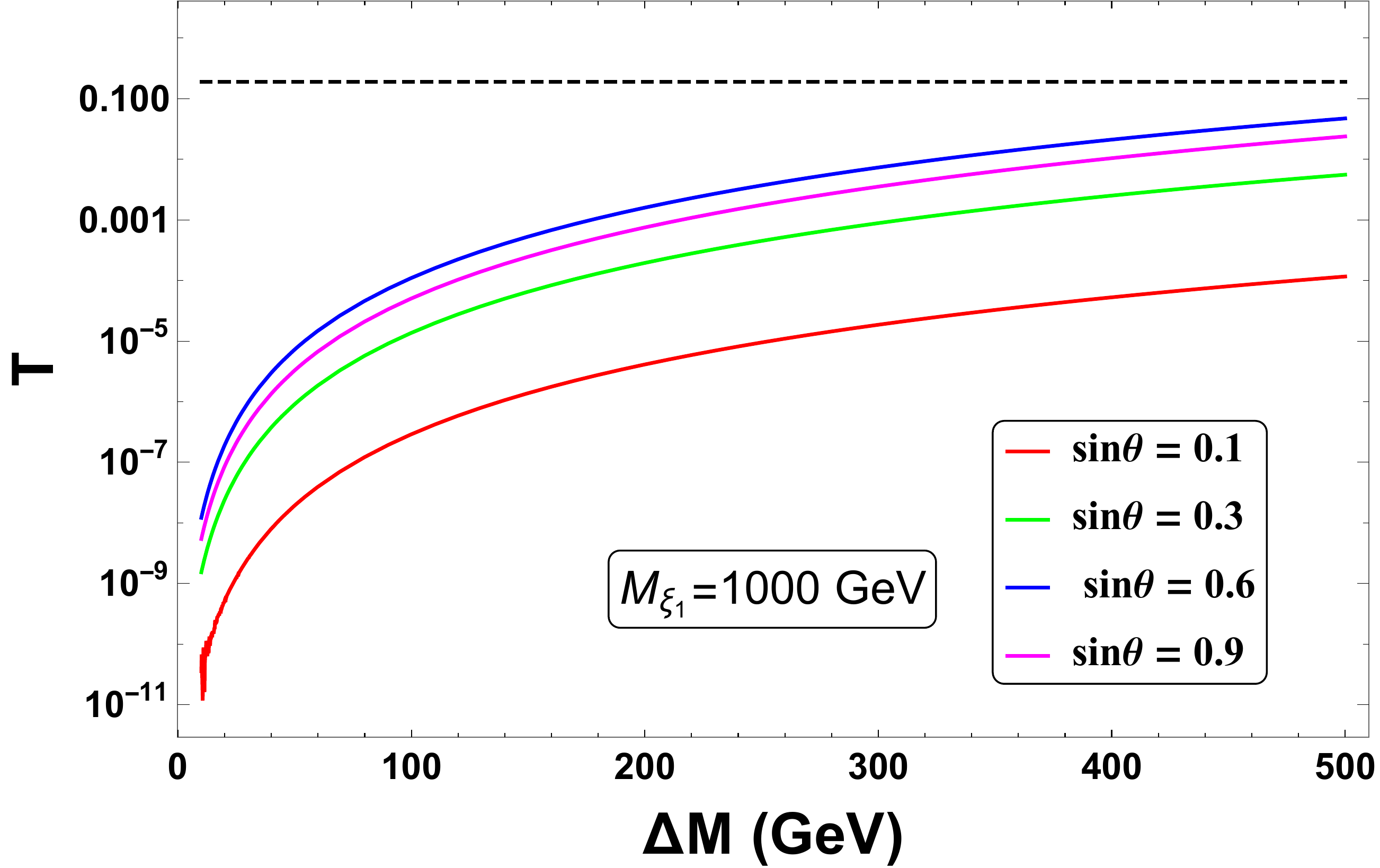}
\caption{Sketch of $T$ parameter using Eq.(\ref{eq:Tpara}) as a function of $\Delta M$ for two different values of $M_{\xi_1}$= 200 GeV (left) and 1000 GeV (right). Each line indicates constant magnitude of $\sin\theta$. The black dashed line stands for the observed upper limit of $T$ parameter. }
\label{fig:Tpara}
\end{figure}

\item \textbf{Electroweak precision observables:} Owing to the presence of an additional $SU(2)_L$ doublet fermion, the electroweak precision parameters put some restrictions on the model parameters.
It turns out that in the small Majorana mass limit the $S$ and $U$ parameters do not pose any significant constraint \cite{Cynolter:2008ea}. However one needs to inspect the magnitude of $T$ parameter originating from the BSM sources. Considering the small Majorana mass limit, the analytical expression for $T$ parameter in our framework carries the following form \cite{Cynolter:2008ea}:
\begin{align}\nonumber
T &~\simeq~ \frac{g^2}{16\pi^2 M_W^2\alpha}\Bigg[\tilde{\Pi}(M_\Psi,M_\Psi,0)+\cos^4\theta \, \tilde{\Pi}(M_{\xi_2},M_{\xi_2},0)+\sin^4\theta \, \tilde{\Pi}(M_{\xi_1},M_{\xi_1},0)\\\label{eq:Tpara}
&+2\sin^2\theta\cos^2\theta \, \tilde{\Pi}(M_{\xi_1},M_{\xi_2},0)-2\cos^2\theta \, \tilde{\Pi}(M_\Psi,M_{\xi_2},0)-2\sin^2\theta \, \tilde{\Pi}(M_{\psi},M_{\xi_1},0)\Bigg]
\end{align}
where $\alpha$ being the fine structure constant. The vacuum polarization functions ($\tilde{\Pi}$) are defined as
\begin{align}\nonumber
\tilde{\Pi}(M_a,M_b)=&-\frac{1}{2}(M_a^2+M_b^2)\Bigg\{\textrm{Div}+{\rm Ln}\left(\frac{\mu^2}{M_a M_b}\right)-\frac{1}{2}\Bigg\}-\frac{(M_a^4+M_b^4)}{4(M_a^2-M_b^2)}{\rm Ln}\left(\frac{M_a^2}{M_b^2}\right)
\\&+M_aM_b\Bigg\{{\rm Div}+{\rm Ln}\left(\frac{\mu^2}{M_aM_b}\right)+1+\frac{M_a^2+M_b^2}{2(M_a^2-M_b^2)}{\rm Ln}\left(\frac{M_b^2}{M_a^2}\right)\Bigg\}.
\end{align}
The present experimental bounds on $T$ is given by \cite{Tanabashi:2018oca}:
\begin{align}
\Delta T=0.07\pm 0.12,
\end{align}

In Fig. \ref{fig:Tpara}, we demonstrate the functional dependence of $T$ parameter on $M_{\xi_1}$, $\Delta M$ and $\sin\theta$. Two notable features come out: (i) for a constant $M_{\xi_1}$ and $\sin\theta$, one can observe the rise of $T$ parameter with $\Delta M$ and thus at some point crosses the allowed experimental upper limit, (ii) for higher DM mass, the constraints on the model variables from $T$ parameter turn weaker. 
\item \textbf{Relic density bound and direct search constraints: }
The observed amount of relic abundance of the dark matter is obtained by the Planck experiment \cite{Aghanim:2018eyx}
\begin{align}
0.1166 \lesssim \Omega_{\rm DM} h^2 \lesssim 0.1206.
\end{align}
Along with this, the dark matter relic density parameter space is constrained significantly by the direct detection
experiments such as LUX \cite{Akerib:2016vxi}, PandaX-II \cite{Cui:2017nnn} and XENON 1T \cite{Aprile:2018dbl}. In our analysis, we will follow the Xenon-1T result in order to validate our model parameter space through direct search bound. 

{\color{black}Here we would like to reinforce the view that although the $Z$-boson mediated spin independent (SI) direct search process can be suppressed at tree level (as commented in the introduction section), the SM Higgs mediated SI direct search process still survives providing loose constraints. Thus the bound on the SI direct search cross section from experiments like Xenon-1T is still applicable. The spin dependent direct search cross section is negligible in our working limit $Y_1\sim Y_2$ (see footnote \ref{ft:axial} for more details).}

\item \textbf{Bounds from invisible decay of Higgs and $Z$ boson:}
In case the DM mass is lighter than half of Higgs or Z Boson mass, decays of Higgs and Z boson to DM are possible. Invisible decay widths of both $H$ and $Z$ are severely restricted at the LHC \cite{Tanabashi:2018oca,Khachatryan:2016whc}, and thus could constrain the relevant parameter space. Since, in the present study our focus would be on the mass range 100 GeV -- 1 TeV for DM, the constraints from $H$ and $Z$ bosons does not stand pertinent.
\end{itemize}

 In the upcoming discussions we will strictly ensure the validity of the above mentioned constraints on the model parameters while specifying the
  the benchmark/reference points that satisfy the other relevant bounds arising from DM phenomenology and leptogenesis.

\section{Fast expanding Universe}\label{fastexpanding}
 As mentioned earlier, the presence of a new species in the early Universe before the radiation domination epoch can significantly escalate the expansion rate of the universe, which in turn has a large impact on the evolution of the particle species present in that epoch. In this section we brief the quantitative justification of the effect of a new species on the expansion rate of the universe. Hubble parameter $H$ delineates the expansion rate of the universe and is connected with the total energy of the Universe through the standard Friedman equation. In presence of a new species ($\eta$) along with the radiation field, the total energy budget of the universe is $\rho = \rho_{\text{rad}} + \rho_\eta$. For standard cosmology, the $\eta$ field would be absent and one can write $\rho=\rho_{\text{rad}}$. As a function of temperature ($T$) one can always express the energy density of the radiation component which is given by
\begin{equation}
 \rho_{\text{rad}} (T) = \frac{\pi^2}{30} g_*(T)T^4,
\end{equation}
with $g_* (T)$ being the effective number of relativistic degrees of freedom at temperature $T$. In the absence of entropy production per comoving volume \textit{i.e.} $sa^3=$ const., one can write $\rho_{\text{rad}}(t)\propto a(t)^{-4} $. Now, in case of a rapid expansion of the Universe the energy density of $\eta$ field is expected to be redshifted quite earlier than the radiation. Accordingly, one can have $\rho_\eta\propto a(t)^{-(4+n)}$ with $n>0$.

The entropy density of the Universe is expressed as $s(T) = \frac{2\pi^2}{45}g_{*s}(T)T^3$ 
where, $g_{*s}$ is the effective relativistic degrees of freedom which contributes to the entropy density. Employing the energy conservation principle once again, a general form of $\rho_\phi$ can thus be constructed as:
\begin{equation}
 \rho_\eta(T) =  \rho_\eta(T_r)\left(\frac{g_{*s}(T)}{g_{*s}(T_r)}\right)^{(4+n)/3}\left(\frac{T}{T_r}\right)^{(4+n)}.
\end{equation}
The temperature $T_r$ is an unknown variable ($ >T_{\text{BBN}}$) and can be safely treated as the point of equality of two respective energy densities: $\rho_\eta(T_r)=\rho_{\text{rad}}(T_r)$. Using this criteria, it is simple to write the total energy density
at any temperature ($T>T_r$) as \cite{DEramo:2017gpl}
\begin{equation}\label{totalrho}
 \rho(T) = \rho_{rad}(T)+\rho_{\eta}(T)=\rho_{rad}(T)\left[1+\frac{g_* (T_r)}{g_* (T)}\left(\frac{g_{*s}(T)}{g_{*s}(T_r)}\right)^{(4+n)/3}\left(\frac{T}{T_r}\right)^n\right]
\end{equation}
From the above equation, it is obvious that the energy density of the Universe at any arbitrary temperature ($T>T_r$), is dominated by $\eta$ component. The standard Friedman equation connecting the Hubble parameter with the energy density of the Universe is given by:
 \begin{equation}
  H = \frac{\sqrt{8\pi\rho}}{\sqrt{3}M_{\text{Pl}}},
 \end{equation}
with $M_{\text{Pl}}= 1.22 \times 10^{19}$ GeV being the Planck mass. At temperature, higher than $T_r$ with the condition $g_*(T) = \bar g_*$ (some \textit{constant}), the Hubble rate can approximately be cast into the following form \cite{DEramo:2017gpl}
\begin{align}\label{modifiedH}
 H(T) &\approx \frac{2\sqrt{2}\pi^{3/2} \bar g_*^{1/2}}{3\sqrt{10}} \frac{T^2}{M_{\text{Pl}}}\left(\frac{T}{T_r}\right)^{n/2}, ~~~~ ({\rm with ~~}T \gg T_r),\\
 &=H_R(T)\left(\frac{T}{T_r}\right)^{n/2}, 
\end{align}
 where $H_R(T)\sim 1.66~\bar{g}_*^{1/2}\frac{T^2}{M_{\rm Pl}}$, the Hubble rate for radiation dominated Universe. In case of SM, $\bar g_*$ can be identified with the total SM degrees of freedom $g_*\text{(SM)} = 106.75$. It is important to note from Eq.(\ref{modifiedH}) that the expansion rate is larger than what it is supposed to be in the standard cosmological background provided, $T>T_r$ and $n>0$. Hence it can be stated that if the DM freezes out during $\eta$ domination, the situation will alter consequently with respect to the one in the standard cosmology.

With positive scalar potential for the field responsible for fast expansion, value of $0< n \leq 2$ can be realized. The candidate for $n=2$ species could be the quintessence fluids \cite{Caldwell:1997ii} where in the kination regime $\rho_\eta \propto a(t)^{-6}$ can be attained. However for $n>2$, one needs to consider negative potential. A specific structure of $n>2$ potential can be found in ref. \cite{DEramo:2017gpl} which is asymptotically free. 

\section{Revisiting dark matter phenomenology}\label{DM}
The comoving number density of the DM ($\zeta_1$) is governed by the Boltzmann's equation (in a radiation dominated Universe) \cite{Kolb:1990vq}:
\begin{align}
 \frac{dY_{\zeta_1}}{dz_D}=-\frac{\langle\sigma v\rangle s}{H_R(T) z_D}(Y_{\zeta_1}^2-Y_{\zeta_1}^{\rm eq^2}),\label{eq:BoltzDM1}
\end{align}
where, $z_D=\frac{M_{\zeta_1}}{T}$ and $\langle\sigma v\rangle$ stands for the thermally averaged annihilation cross section with $v $ being the relative velocity of the annihilating particles. The equilibrium number density of the DM component is represented by $Y_{\zeta_1}^{\rm eq}$ in Eq.(\ref{eq:BoltzDM1}). The relic abundance of the DM is obtained by using \cite{Kolb:1990vq}:
\begin{align}
\Omega_{\rm DM} h^2= 2.82\times 10^8~  M_{\zeta_1} Y_{z_D=\infty}
\end{align}
In the WIMP paradigm, it is presumed that  DM stays in thermal equilibrium in the early Universe. Considering the DM freezes out in the RD Universe, the required order of thermally averaged interaction strength of the DM to account for correct relic abundance is found to be,
\begin{equation}\label{relicab}
 \langle \sigma v \rangle \approx 3 \times 10^{-26}\text{cm}^3 ~\text{sec}^{-1},
\end{equation}
The Eq.(\ref{relicab}) quantifies an important benchmark for WIMP search, which bargains on a major assumption that the universe was radiation dominated at the time of DM freeze out. However, in an alternative cosmological history, depending on the decoupling point of DM from the thermal bath this number is expected to change by order of magnitudes, which in turn, brings out significant changes in the  relic satisfied parameter space of a particular framework. 

In the current framework, the DM $\zeta_1$ can (co-)annihilate with the {\color{black} other heavier neutral and charged fermions into SM particles through $Z$ or Higgs mediation.} Furthermore, co-annihilation processes like $\psi^+\psi^-\rightarrow$~SM,~SM ({ $\psi^\pm$ are the charged counterpart of the vector fermion doublet $\Psi$}) also supply their individual contributions to total $\langle\sigma v\rangle$. The relevant Feynman diagrams contributing to the possible annihilation and co-annihilation channels of the DM can be found in \cite{DuttaBanik:2018emv}. { For the model implementation we have used \textsf{Feynrules} \cite{Alloul:2013bka} and subsequently \textsf{Micromega} \cite{Belanger:2014vza} to carry out the DM phenomenology.}

As mentioned in the previous section for the fast expanding Universe the Hubble parameter $H_R(T)$ in Eq.(\ref{eq:BoltzDM1}) in presence of the new species $\eta$, need to be replaced with $H(T)$ of Eq.(\ref{modifiedH}) with $n>0$. This recent temperature dependence of the expansion rate of the Universe provide some new degrees of freedom as we also observe here. For the standard cosmological background, in pseudo Dirac singlet doublet dark matter model there are three independent parameters for a particular DM mass namely: $\Delta M,~\sin\theta$ and the Majorana mass $m$\footnote{\label{ft:axial}\color{black}It is worth mentioning that for a general case where the two Yukawas are not equal, one has to deal with two mixing angles, namely $\theta_L$ and $\theta_R$ rather considering only one ($\theta$).  In the pseudo Dirac case with $\theta_L\neq \theta_R$ (or $Y_1\neq Y_2)$, a few extra axial type interactions for DM ($\zeta_1$) appear in the Lagrangian which vanish in the $\theta_L\sim\theta_R$ (or $Y_1\sim Y_2$) limit. These axial couplings have negligible contribution to the DM relic abundance as we have checked. Having said that, one of the axial interactions of DM $\sim \overline{\zeta_1}\gamma_\mu\gamma_5\zeta_1Z^\mu$ (with coupling coefficient proportional to $ \sin^2 \theta_R - \sin^2 \theta_L$) can yield non zero spin dependent nucleon cross section for $\theta_L\neq\theta_R$ which can provide signal in the spin dependent direct search experiments. Since, one of our major aims of the present study is to hide the DM at both spin independent and spin dependent direct search experiments, we work with the pseudo-Dirac and $\theta_L\sim\theta_R$ limits respectively. This further simplifies the scenario, with a single Yukawa like coupling in the set up which is sufficient to portray the novel features of the proposed scenario.}. For simplicity of our analysis we keep the Majorana mass $m$ small by fixing it at 1 GeV. 
 Then the relevant set of parameters which participate in the DM phenomenology in presence of the modified cosmology are the following :
\begin{align}
\Big\{\Delta M, ~\sin\theta,~T_r,~n \Big\},
\end{align}
for a certain DM mass.

\subsection{Spin independent direct search}
\begin{figure}[t]
\centering
\includegraphics[height=3.7cm,width=4cm]{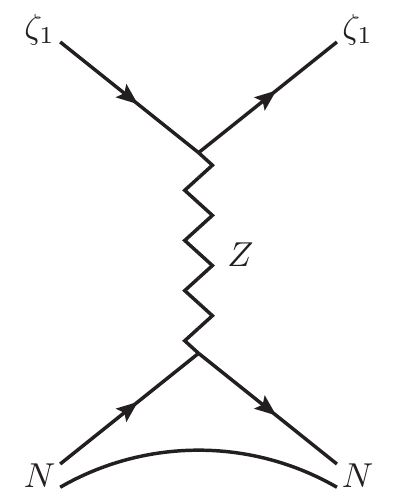}~~~~~~~~~
\includegraphics[height=3.7cm,width=4cm]{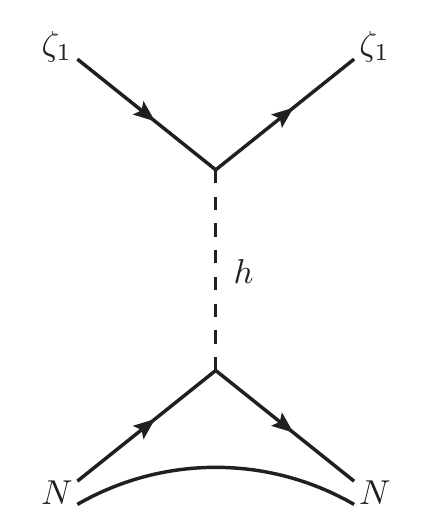}
\caption{Feynman diagrams contributing to the {\color{black}spin independent }direct search of the DM.}
\label{fig:DirectD} 
\end{figure}

The part of the Lagrangian relevant for {\color{black}spin independent} direct search of the DM within the Dirac limit ($m\rightarrow 0$) is given by, 
\begin{align}\label{eq:DD1}
 \mathcal{L}\supset \frac{g}{2 \cos\theta_W} \sin^2\theta~\overline{\xi_1}\gamma^\mu Z_\mu\xi_1+\frac{Y}{\sqrt{2}}\sin\theta\cos\theta~ h~\bar{\xi_1}\xi_1,
\end{align}
However switching the parameter $m$ on, leads to the pseudo-Dirac limit in which the neutral current interaction of the DM $\zeta_1$, i.e., first term of Eq.(\ref{eq:DD1}) vanishes at zeroth order in $\delta_r=\frac{m_{\chi_L}-m_{\chi_R}}{M_{\zeta_1}}$. Although a small residual vector-vector interaction of the DM to the quarks, due to the non-pure Majorana nature of the mass eigenstates still exists {\color{black}at leading order in $\delta_r$}. This brings about the $Z$ mediated effective interactions of the DM with nucleon which is given by,
 \begin{align}
\mathcal{L}\supset\alpha~\delta_r~(\bar{\zeta_1}\gamma^\mu\zeta_1)(\bar{q}\gamma_\mu q),
\end{align}
with $\alpha=\left(\frac{4 g^2\sin^2\theta }{ m_Z^2\cos^2\theta_W}\right)C_V^q=\alpha^\prime C_V^q$ and $g$ as the $SU(2)_L$ gauge coupling constant. In addition, the SM Higgs mediated process of DM-nucleon scattering will be present at the tree level as evident from Eq.(\ref{eq:DD1}). The relevant Feynman diagrams are shown in Fig. \ref{fig:DirectD}. It is pertinent to comment that in the vanishing $\delta_r$ limit only Higgs mediated diagram in Fig. \ref{fig:DirectD} contribute to the SI direct search of DM. 

\subsection{Dark matter in presence of (non)standard thermal history} 
In case of a faster expansion of the Universe, the DM freezing takes place quite earlier than what it does in the standard scenario, resulting into an overabundance. Hence, to account for the observed relic abundance, an increase of the total annihilation cross section of DM is required. This in turn necessitates the rise of the associated coupling coefficients. 

\begin{figure}[t]
\hspace{-5mm}
\includegraphics[height=6cm,width=7.5cm]{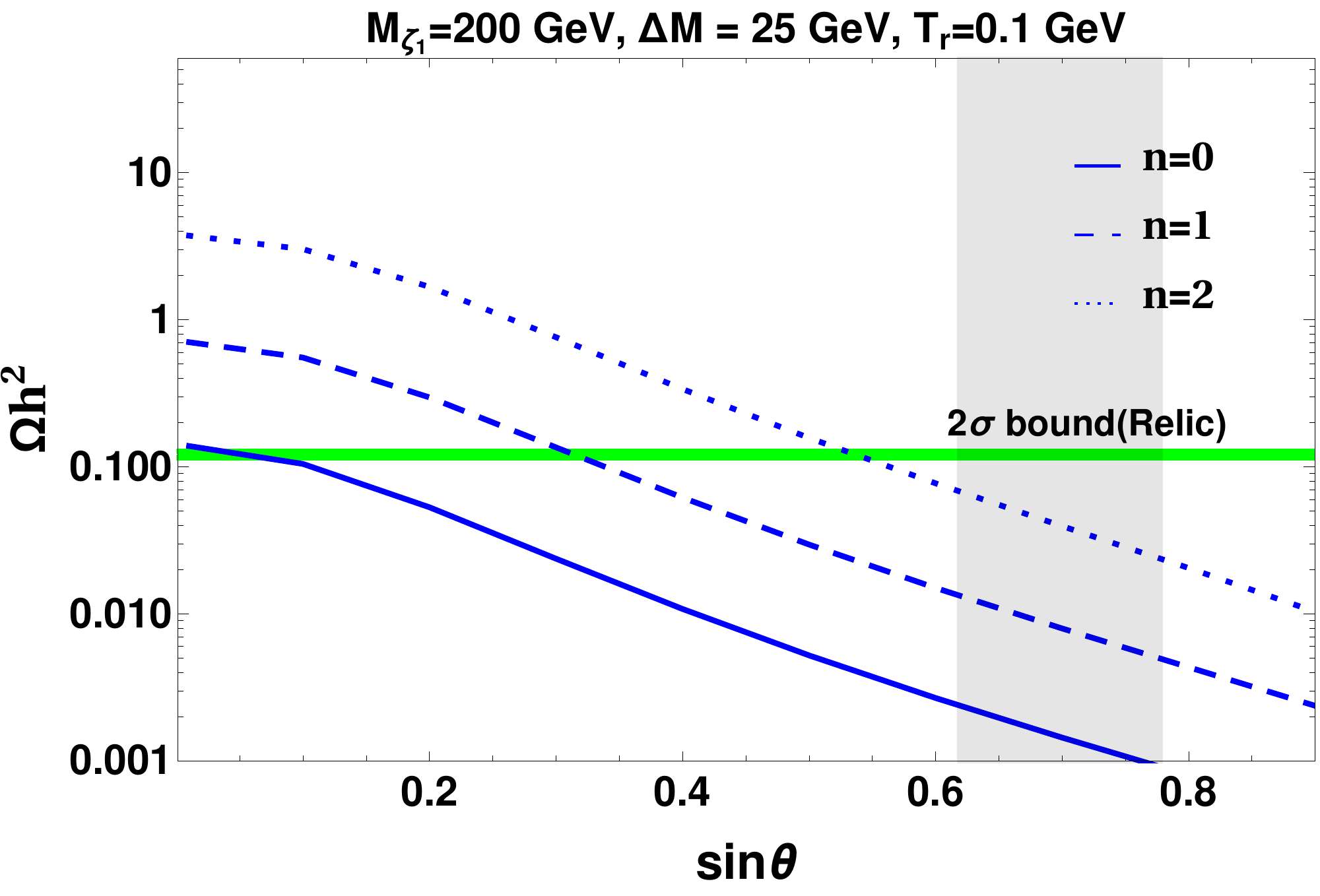}~~ 
\includegraphics[height=6cm,width=7.51cm]{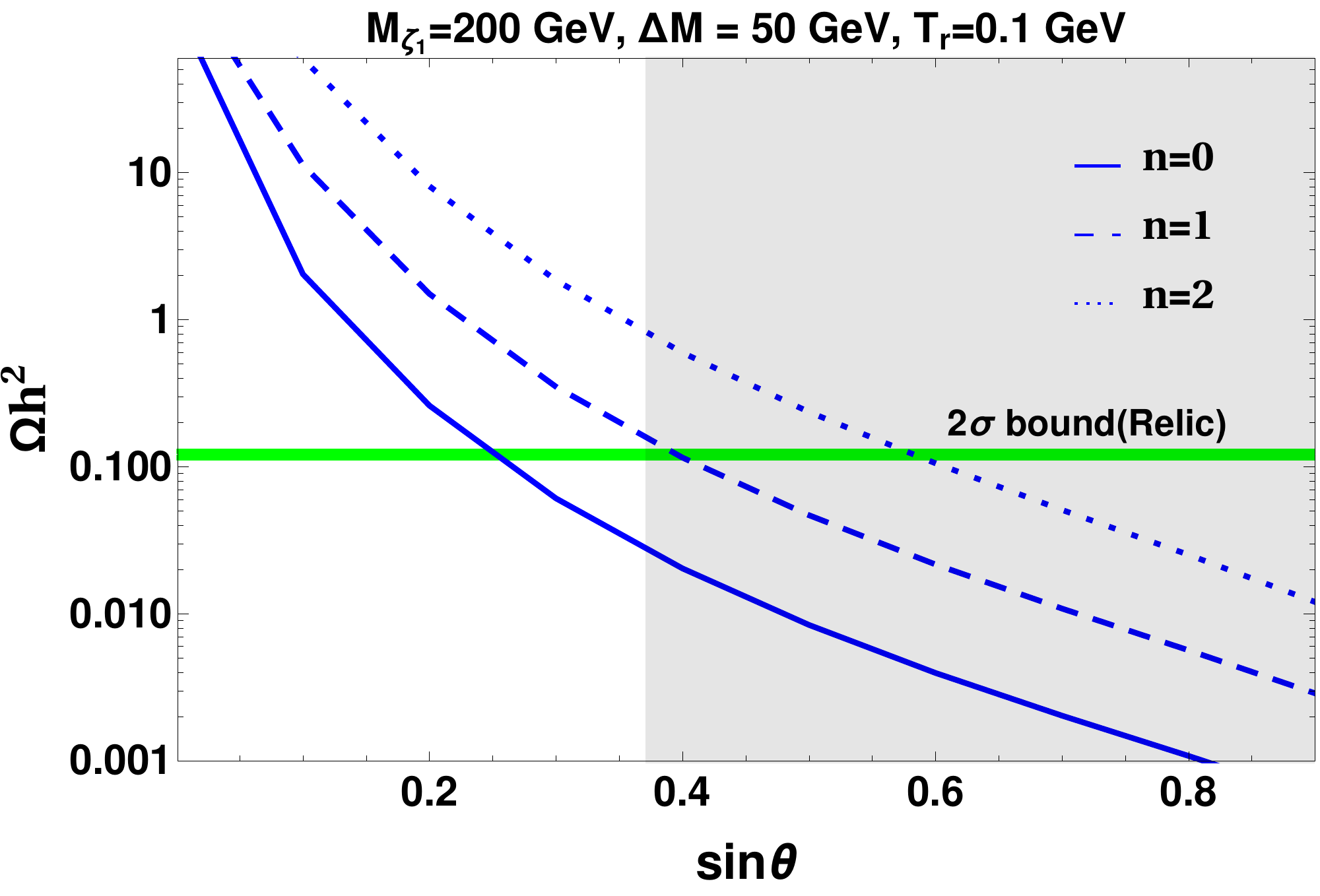}
\caption{Relic abundance of the DM as a function of the mixing angle between the singlet and doublet is shown considering both standard (solid line) and non-standard (dashed and dotted lines) thermal history of the Universe, for $M_{\zeta_1}=200$ GeV with (left) $\Delta M=25$ GeV and (right) $\Delta M=50$ GeV. The disfavored region from the spin independent direct detection constraints are denoted by respective shaded region. Here we have considered $T_r= 0.1$  GeV.}
\label{relicsintheta2} 
\end{figure} 

\begin{figure}[]
\hspace{-5mm}
\includegraphics[height=6cm,width=7.5cm]{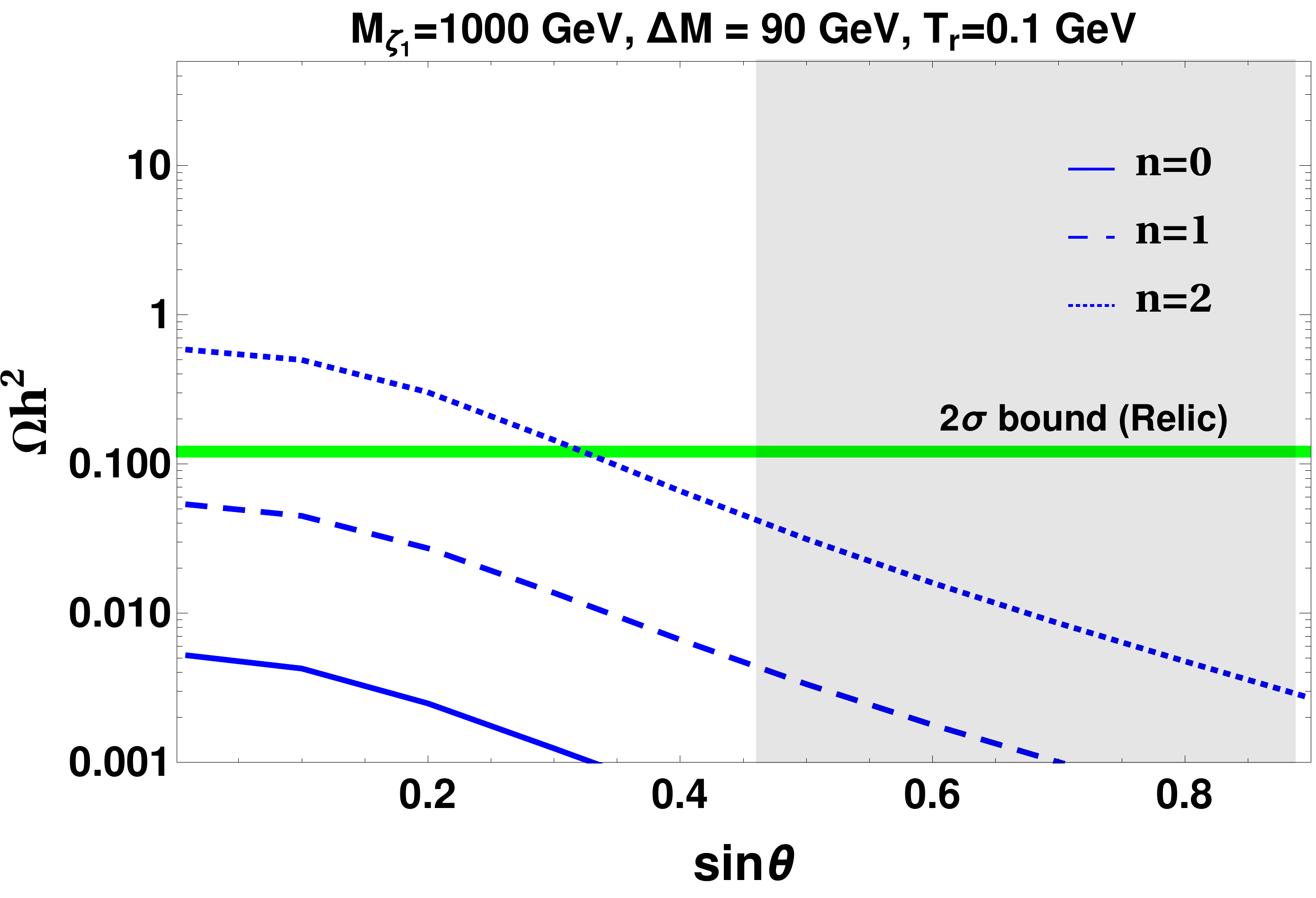}~~ 
\includegraphics[height=6cm,width=7.5cm]{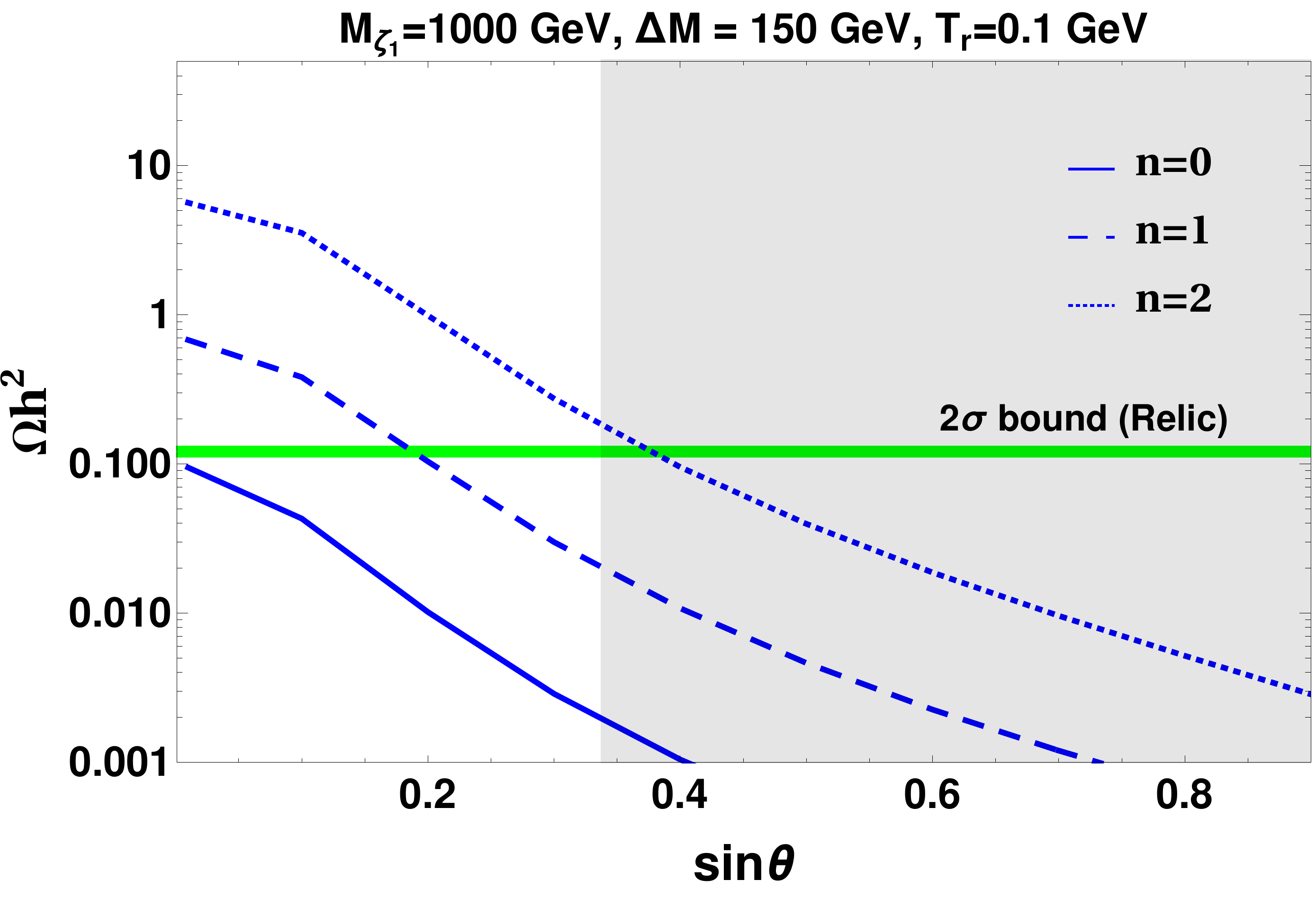}
\caption{The same as Fig. \ref{relicsintheta2}  but for a choice of higher DM mass, for $M_{\zeta_1}=1000$ GeV with (left) $\Delta M=90$ GeV and (right) $\Delta M=150$ GeV.  Here we have fixed $T_r= 0.1$ GeV.}
\label{relicsintheta1} 
\end{figure}

This fact can be realized from Figs. \ref{relicsintheta2}-\ref{relicsintheta1}, where the DM relic abundance is plotted against $\sin\theta$ by considering $T_r=0.1$ GeV. We choose two different DM masses for the analysis, one at a comparatively lower range with $M_{\zeta_1}=200$ GeV shown in Fig. \ref{relicsintheta2} while the other one in a higher mass regime at $M_{\zeta_1}=1000$ GeV as in Fig. \ref{relicsintheta1}. We also take different values of $n$ and $\Delta M$ to have a clear comprehension of how the new degrees of freedom changes the relic density. It is prominent that a larger value of $\sin\theta$ is required in order to satisfy the observed density limit (green color band representing $2\sigma$ range of the observed relic density) for $n\gtrsim 1$ compared to the $n=0$ (standard) case. We also display the SI direct search constraints on the same plot. The contribution to spin independent direct detection cross section comes solely from the Higgs mediated diagrams (right panel of Fig. \ref{fig:DirectD}) since we are working in the $\delta_r=0$ limit. The direct detection cross section seemingly restricts the value of $\sin\theta$ in an intermediate range.  For example, such a constraint of $0.62\lesssim \sin\theta\lesssim 0.76$ is indicated as shaded region in left panel of Fig. \ref{relicsintheta2}. This is because the SI direct search cross section is proportional to the factor: $\sin^2\theta\cos^2\theta$, as evident from Eq.(\ref{eq:DD1}). In the right panel of the Figs. \ref{relicsintheta2}-\ref{relicsintheta1}, this intermediate range (specifically the upper limit) of $\sin\theta$ is not apparently visible since it exceeds the plotting range.

\begin{figure}[t]
\hspace{-5mm}
\includegraphics[height=6cm,width=7.5cm]{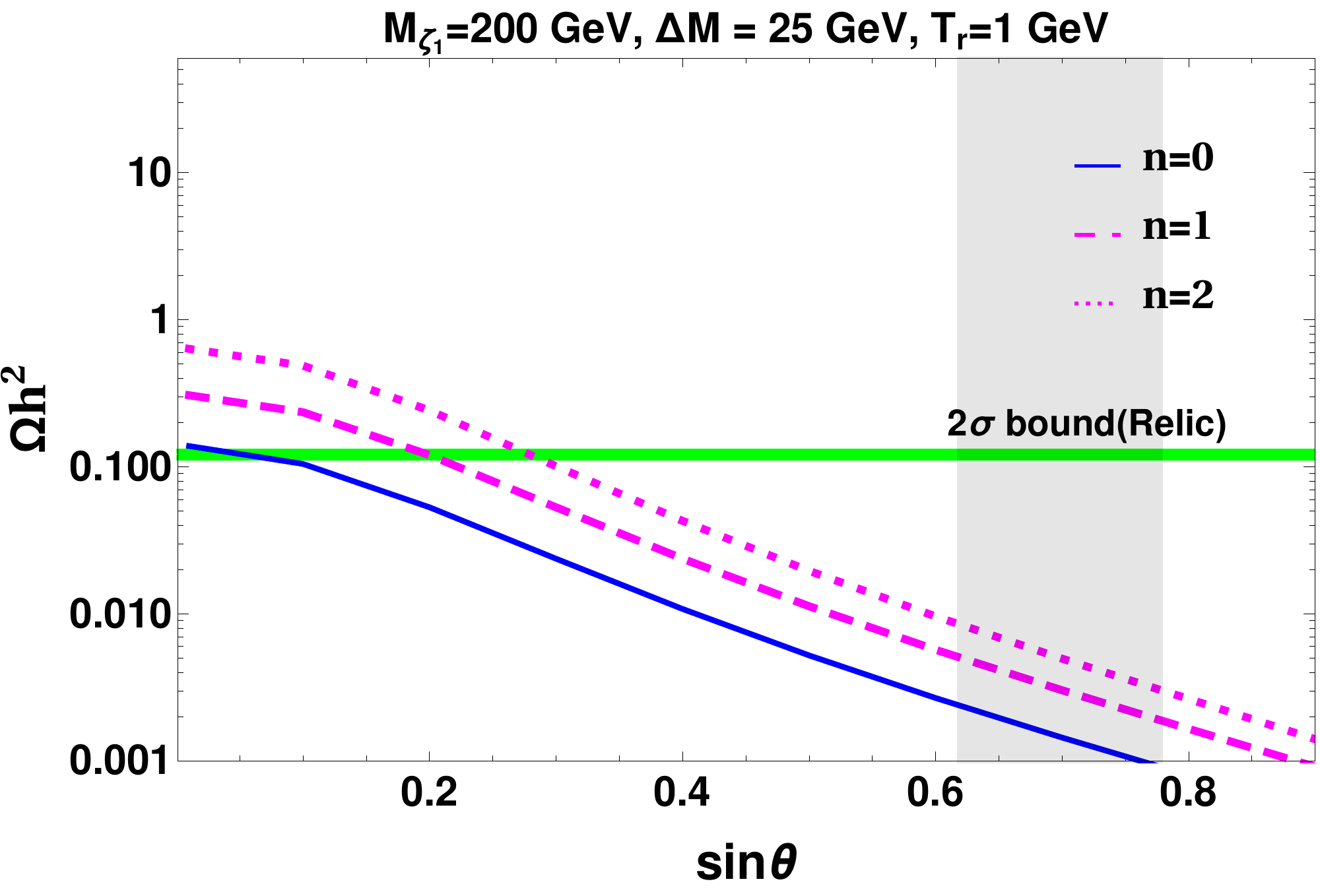}~~ 
\includegraphics[height=6cm,width=7.5cm]{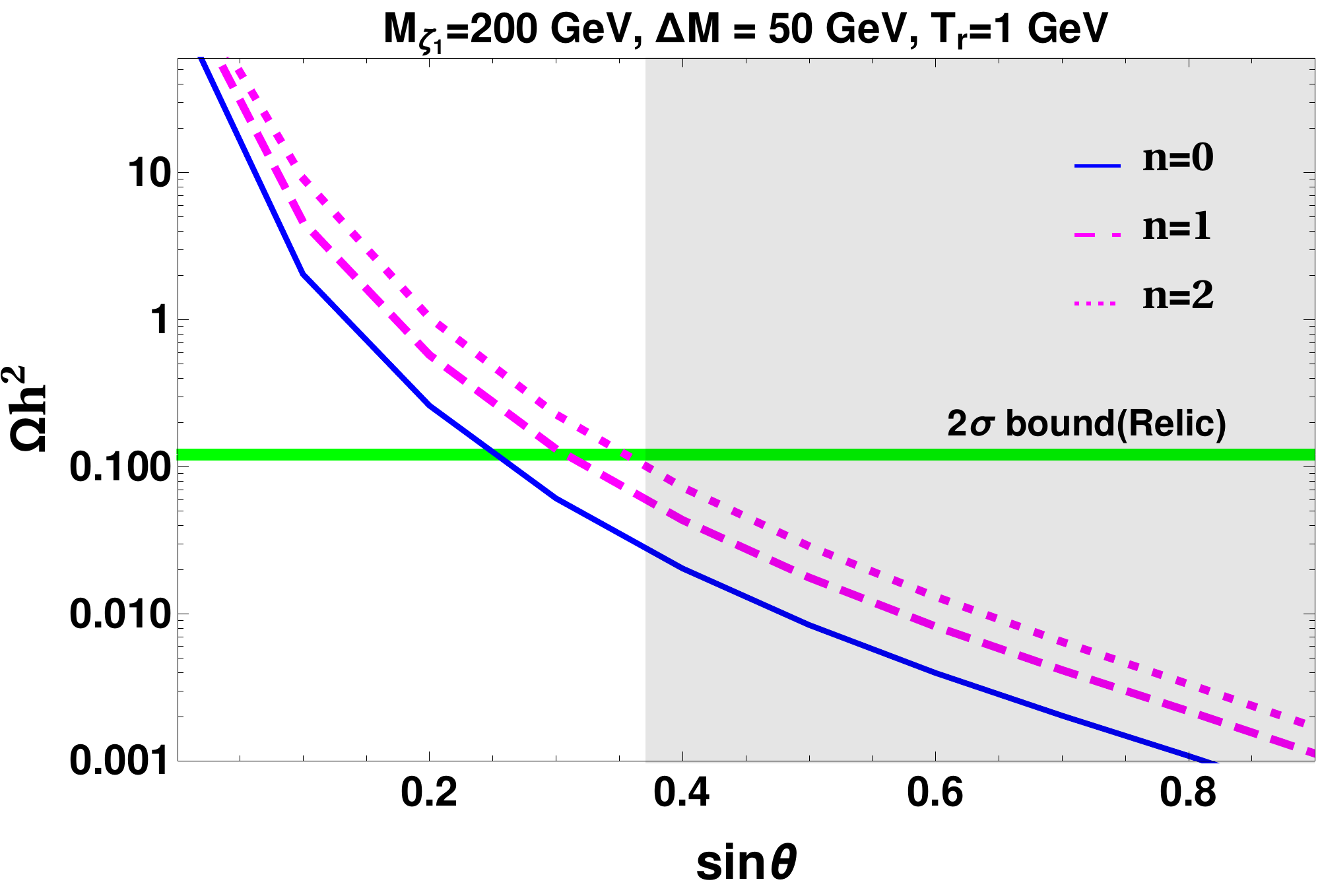}
\caption{The same as Fig. \ref{relicsintheta2}  but for a choice of larger $T_r= 1$ GeV.}
\label{relicsintheta3} 
\end{figure}

A few important aspects of the analysis can be drawn from Figs. \ref{relicsintheta2}-\ref{relicsintheta1}. It is seen that for a particular DM mass, non-standard cosmology ($n>0$) requires larger $\sin\theta$ to be consistent with the observed relic abundance as mentioned earlier. For a specific value of $n$, relic density increases with $\Delta M$ thus at some point can be ruled out from SI direct search bound for a specific DM mass. For example, in the left panel of Fig. \ref{relicsintheta2}, fixing $n=2$ can satisfy the correct relic and which is also allowed by the SI direct search bound. However once $\Delta M$ is increased up to a substantial amount it enters into the disfavored region, as seen in the right panel of Fig. \ref{relicsintheta2}. 

\begin{table}[b]
\begin{center}
\begin{tabular}{|c| c | c | c | c | c | c | c |}
  \hline
  BP & $~n~$ & $T_r$ (GeV) & $M_{\zeta_1} $ (GeV) & $\Delta M$ (GeV) & $\sin\theta$ &  $\Omega h^2$ & $ {\rm Log}_{10}\left[\frac{\sigma ^{\rm SI}}{\text{cm}^2}\right]$  \\
  \hline
  I & 2 & 0.1 & 200 &  25  & 0.53  &0.12 & -46.71    \\
  \hline
  II & 2 & 0.1 &1000 &  90  & 0.325 & 0.12 & -46.8 \\
  \hline
\end{tabular}
\caption{Two sets of relic and SI direct search satisfied points collected from Figs. \ref{relicsintheta2}-\ref{relicsintheta1}}.
\label{tab:tab2}
\end{center} 
\end{table}

So far the DM phenomenology has been studied by assuming $T_r=0.1$ GeV. Nonetheless one can look for the DM parameter space considering a higher value of $T_r$. In Fig. \ref{relicsintheta3}, we use a slightly larger value of $T_r=1$ GeV and present the relic contours for different values of $n$ in $\Omega h^2-\sin\theta$ plane. It is observed that increase of $T_r$ reduces the relic density for a particular $n$. As an example, in the left panel of Fig. \ref{relicsintheta2}, the required value of $\sin\theta$ was 0.53 to satisfy the relic abundance criteria considering $n=2$ and $T_r=0.1$ GeV. Now for $T_r=1$ GeV, this value got shifted to 0.25. Enhancement of $T_r$ is also preferred in the view of SI direct search constraints as can be seen by comparing the right panel of Fig. \ref{relicsintheta2} and Fig. \ref{relicsintheta3} where the $n=2$ relic contour turns out to be favored in the later case. This leads to a realization that, lowering the required value of $\sin\theta$ to account for the expected relic density further reduces the SI direct search cross section. One can assign a further higher value to $T_r>1$, however the scenario will approach towards the standard case which is prominent in comparing Fig. \ref{relicsintheta2} and Fig. \ref{relicsintheta3}. 
We end this section by tabulating two sets of relic satisfied points for $n=2$ in Table \ref{tab:tab2} which have relevance in the study of neutrino mass and leptogenesis.

\section{Neutrino mass generation}\label{sec:neu}

\begin{figure}
\centering
 \includegraphics{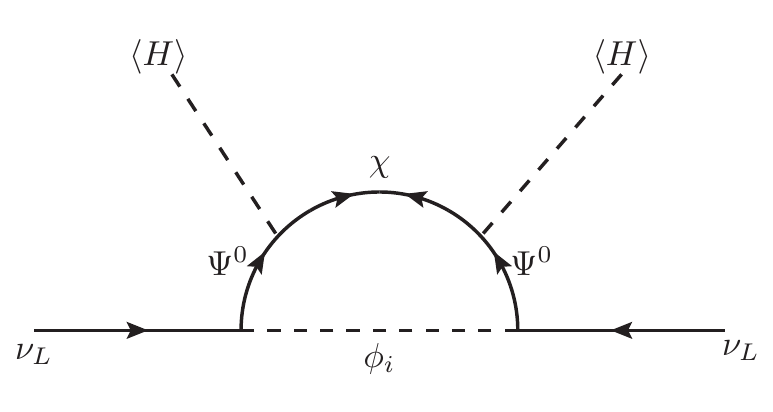}
 \caption{Schematic diagram of radiative neutrino mass generation.}
 \label{neutrinomass}
\end{figure}

This model renders a mechanism which explains the radiative generation of light neutrino mass. The relevant one loop process is shown in Fig. \ref{neutrinomass} which establishes the fact that the presence of the heavy scalars are essential in order to make the Majorana light neutrinos massive.
The light neutrino mass matrix can be expressed by the following equation \cite{Ma:2006km,Ma:2009gu,Fraser:2014yha}:
\begin{align}\label{mnu}
m_{\nu_{\alpha \beta}}=h^T_{i\alpha} \Lambda_{ii} h_{\beta i}, 
\end{align}
where, 
$\Lambda_{ii}=\Lambda_{ii}^L+\Lambda_{ii}^R$. The $\Lambda_{ii}^L$ and $\Lambda_{ii}^R$ include the contribution from $m_{\chi_L}$ and $m_{\chi_R}$ respectively. For the full analytical expressions representing $\Lambda_{ii}^L ,~ \Lambda_{ii}^R$ we refer to our earlier work \cite{Konar:2020wvl}. We use Casas-Ibarra parameterization ~\cite{Casas:2001sr} in order to connect the mixing parameters with neutrino Yukawa coupling. Using this parameterization, one can write \cite{Casas:2001sr}:
\begin{equation}\label{eq:casas}
  h^T = D_{\!\!\sqrt{\Lambda^{-1}}} \, \mathcal{R} \, D_{\!\!\!\!\sqrt{m_\nu^{\rm diag}}} \, U^{\dagger},
 \end{equation}
 where, $\mathcal{R}$ is a complex orthogonal matrix. Any complex orthogonal matrix can be manifested by $\mathcal{R}= O ~ e^{i A}$ where $O$ and $A$ represent any arbitrary real orthogonal and real anti-symmetric matrices respectively \cite{Pascoli:2003rq}. The exponential of the anti-symmetric matrix $A$ can be simplified to
  \begin{equation}
  e^{i A} = 1- \frac{\cosh r -1}{r^2}A^2 + i \frac{\sinh r}{r} A
  \end{equation}
  with $r = \sqrt{a^2+b^2+c^2}$ and
 \begin{equation}
 A = \left(\begin{matrix}
 0 & a & b\\
 -a & 0 & c\\
 -b & -c & 0
 \end{matrix}\right)
 \end{equation}
 
For our purpose, we consider $O$ as an identity matrix and also for simplicity of the anti-symmetric matrix $A$ we have
chosen the equality $a= b = c \equiv a$. It is important to note that, this particular parameterization for 
 the $\mathcal{R}$ matrix helps us to achieve a desired order of Yukawa coupling by keeping the neutrino mixing parameters intact.
 We denote, $D_{\!\!\!\!\sqrt{m_\nu^{\rm diag}}} = {\rm Diag}(\sqrt{m_{\nu 1}},~\sqrt{m_{\nu_ 2}},~\sqrt{m_{\nu 3}}), ~~ D_{\!\!\sqrt{\Lambda^{-1}}} = {\rm Diag}(\sqrt{\Lambda_{11}^{-1}},~\sqrt{\Lambda_{22}^{-1}},~\sqrt{\Lambda_{33}^{-1}})$.  It is also worth mentioning that this special kind of Casas-Ibarra parametrization for the neutrino Yukawa coupling is found to be facilitating to produce the parameter space responsible for generating the observed BAU in the present framework. 
 Authors in \cite{Petcov:2006pc} have shown the explicit roles of the anti-symmetric matrix $A$ and its elements $a,b,c$ in order to achieve sufficient amount of lepton asymmetry. In our case too, the usefulness of this particular parametrization can be observed in Section \ref{resultsL} where we tune $a$ such that one can acquire the observed BAU. 
 As obtained from the recent bayesian analysis \cite{Capozzi:2018ubv}, the mild preference for the normal mass hierarchy (NH) of the neutrinos, allows us to chose the NH as the true hierarchy among the three light neutrino masses. It is also found that the latest global fit of neutrino oscillation data \cite{Esteban:2018azc}  seems to favor the second octant of the atmospheric mixing angle for both the mass hierarchies. The recent announcement made by the experiment prefers the Dirac CP phase to be  $-\pi/2$ with $3\sigma$ confidence level (for detail one may refer to \cite{Abe:2019vii}). Keeping all these in mind for the numerical analysis section we fix all the neutrino parameters to their $3\sigma$ central values including the maximal values for the Dirac CP phase. It is also noted that, a random scan of all the neutrino parameters in their entire $3\sigma$ range would not affect our present analysis much. The resulting Yukawa coupling in the neutrino sector governs the CP violating decay of the BSM scalar leading to an expected amount of lepton asymmetry which we discuss in the next section.
\section{Baryogenesis via Leptogenesis from scalar decay}\label{leptogenesis}
In this section, we describe the production mechanism of lepton asymmetry driven by the decay of the scalar belonging to the dark sector. Our proposal for leptogenesis differs from the usual scenario of leptogenesis in the type I seesaw framework in the sense that, in such a scheme the production of lepton asymmetry is guided by
the decay of the heavy Majorana RHN. The present set up, on account of the presence of lepton number violating vertex involving $\phi$ and the SM leptons, motivates us to investigate the process of lepton asymmetry creation from the singlet scalar ($\phi$) decay which has also served a key role in generating the light neutrino mass. We will also see that presence of a non-standard history of the early Universe provides indisputable contribution in order to yield correct order of baryon asymmetry by suppressing the washout factor significantly.

In the present framework the dark sector scalar ($\phi$) can undergo a CP violating decay to SM leptons and the additional BSM fermion doublet which
leads to lepton number violation by one unit. This particular decay process can naturally create lepton number asymmetry provided out-of-equilibrium criteria is satisfied. Earlier we have commented on the choice of the mass spectrum of dark sector scalars \textit{i.e.} $M_{\phi_{1}} < M_{\phi_{2}} < M_{\phi_{3}} $ (see Fig. \ref{massorder}), which however do not play any decisive role in favoring the true hierarchy of neutrino mass. All these scalars can potentially contribute to generate the final $B-L$ asymmetry. 
The CP asymmetry factor is defined as the ratio of the difference between the decay rates of $\phi$ into the final state particles with lepton number +1 and -1 to the sum of all the
decay rates,  quantified as,
\begin{equation}\label{eq:assym1}
 \epsilon_i^\alpha = \frac{\Gamma(\phi_i \rightarrow \bar L_\alpha \Psi) - \Gamma(\phi_i \rightarrow  L_\alpha \bar \Psi)}{\Gamma(\phi_i \rightarrow \bar L_\alpha \Psi) + \Gamma(\phi_i \rightarrow  L_\alpha \bar \Psi)},
\end{equation}
\begin{figure}[t]
\centering
\includegraphics[height=2.3cm,width=4.2cm]{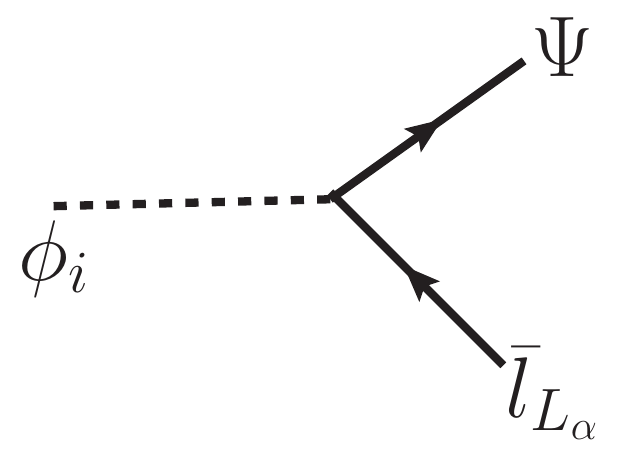}~
\includegraphics[height=2.7cm,width=5cm]{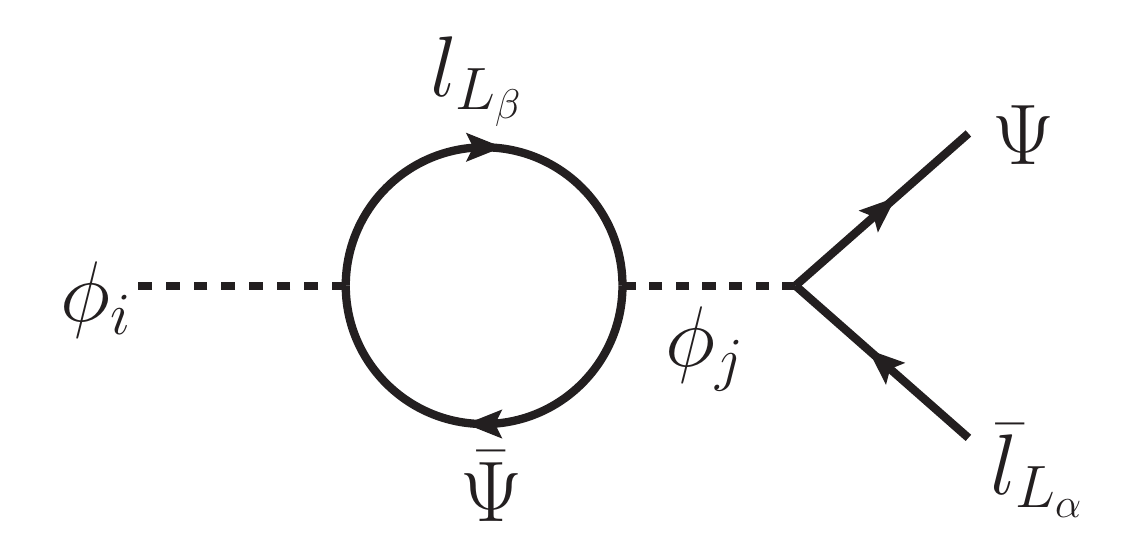}~
\includegraphics[height=2.3cm,width=4.2cm]{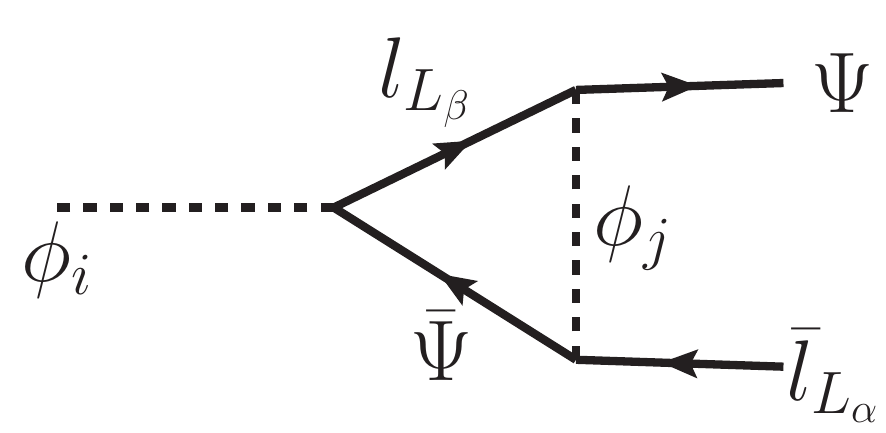}
\caption{Possible feynman diagrams for lepton asymmetry production from singlet scalar decay}
\label{feyn} 
\end{figure}
\noindent The total lepton asymmetry receives contributions from two kind of subprocesses: (i) superposition of tree level and vertex diagram and (ii) superposition between tree level and self energy diagram  
as shown in Fig \ref{feyn}. This allows us to write $\epsilon_T = \epsilon_{\text{vertex}} + \epsilon_{\text {self energy}}$. Driven by Eq.(\ref{eq:assym1}), we can obtain the analytical form of $\epsilon_{\text{vertex}}$ which is given by (see Appendix \ref{appen1} for the detail):
\begin{equation}\label{eq:ep1}
 \epsilon_{\text{vertex}}^i = \frac{1}{4\pi}\sum_{j\neq i}\frac{\text{Im} \left[(h^\dagger h)_{ij}h_{\alpha j}h_{\alpha i}^*\right]}{(h^\dagger h)_{ii}}x_{ij} \text{~log}\left(\frac{x_{ij}}{x_{ij}+1}\right)
\end{equation}
where, $h_{\alpha i}$ is the Yukawa matrix governing the lepton number violating interaction in this set up and $x_{ij} =\frac{M_{\phi_j}^2}{M_{\phi_i}^2}$. In computing Eq.(\ref{eq:ep1}) we have considered the massless limit for the SM leptons. We also have figured out that the $\epsilon^i_{\text {self energy}}$ exactly vanishes in this limit. 
A more detailed analytical understanding of this asymmetry parameter is provided in the Appendix
\ref{appen1}. The obtained amount of lepton asymmetry can estimate the observed BAU in presence of a rapid expansion of the Universe for a particular domain of scalar mass. The effect of this unorthodox cosmology is crucial especially in bringing down the leptogenesis scale and can be realized from the modifications brought out in the Boltzmann's Equations which we are going to discuss in the following subsection.

\subsection{Boltzmann's equations and final baryon asymmetry}
 The evolutions of number densities of $\phi$ and $B-L$ asymmetry can be obtained by solving the following set of coupled Boltzmann's equations~(BEQs) \cite{Buchmuller:2004nz,Davidson:2008bu}:
\begin{gather}\label{eq:BeL}
 \frac{dN_{\phi_{i}}}{dz} = - D_i (N_{\phi_{i}} - N^{eq}_{\phi_{i}}), ~~~ \text{with}~~~ i = 1,2,3\\ 
 \frac{dN_{B - L}}{dz} = -\sum_{i = 1}^3 \epsilon_i D_i (N_{\phi_{i}} - N^{eq}_{\phi_{i}})-\sum_{i = 1}^3 W_i N_{B - L},
\end{gather}
with $z = M_{\phi_1} /T$ when the decaying scalar is the $\phi_1$. For convenience in numerical evaluation in case all the three scalars are actively involved in the generation of the final lepton asymmetry (which is true here) one can redefine a generalized temperature-function ($z$),  writing $z=\frac{z_i}{\sqrt{x_{1i}}}$ with $i = 1,2,3$. Note that $N_{\phi_i}$'s are the comoving number densities normalised by the photon density at temperature larger than $M_{\phi_i}$. The first one of the above set of coupled equations tells us about the evolution of the scalar number density whereas the second determines the evolution of the amount of the lepton asymmetry which survives in the interplay of the production from parent particle (first term) and washout (second term), as a function of temperature. 

To properly deal with the wash out of the produced lepton asymmetry one must take into account all the possible processes which can potentially erase a previously created asymmetry.
Ideally there exist four kinds of processes which contribute to the different terms in the above BEQs: decays, inverse decays, $\Delta L = 1$ and $\Delta L =2$ scatterings mediated by the decaying particle.  In the weak washout regime,  the later two processes contribute negligibly to the washout.  Hence in our present analysis, considering an initial equilibrium abundance\footnote{In the case of vanishing initial abundance of $N_1$, the $\Delta L = 1$ scatterings can enhance the abundance of $N_1$ and increase the  efficiency factor \cite{Plumacher:1996kc,Buchmuller:2004tu}} of $N_1$, the inverse decay offers the principal contribution.

The Hubble expansion rate in the standard cosmology is estimated to be $H_R(T) \approx \sqrt{\frac{8 \pi^3 g_*}{90}}\frac{M_{\phi_1}^2}{M_\text{Pl}}\frac{1}{z^2}\approx 1.66 g_*\frac{M_{\phi_1}^2}{M_{\text{Pl}}}\frac{1}{z^2}$ with $g_* = 106.75$, being the effective relativistic degrees of freedom.
 The $D_i$ in Eq.(\ref{eq:BeL}) denotes the decay term which can be expressed as,
\begin{equation}\label{eq:ratio1}
 D_i = \frac{\Gamma_{D, i}}{H z} = K_i x_{1i}z \langle 1/ \gamma_i\rangle,
 \end{equation}
considering $H=H_R$ and one can write $\Gamma_{D, i} = \bar \Gamma_i + \Gamma_i = \tilde{\Gamma}_{D,i}\langle 1/ \gamma_i\rangle$ with $\langle 1/ \gamma_i\rangle$, the ratio of the modified Bessel functions $\mathcal{K}_1$ and $\mathcal{K}_2$ quantifying the thermally averaged dilution factor as $ \langle 1/ \gamma_i\rangle = \frac{\mathcal{K}_1(z_i)}{\mathcal{K}_2(z_i)}$. Note that $\Gamma_i$ represents the thermally averaged decay width of $\phi_i$ to SM lepton and the BSM fermion doublet whereas $\bar{\Gamma}_i$ stands for the conjugate process of the former. The wash out factor $K_i$ in Eq.(\ref{eq:ratio1}) is related to the decay width and the Hubble expansion rate as
\begin{align}
 K_i \equiv \frac{\tilde{\Gamma}_{D,i}}{ H (T=M_{\phi_i})}.
 \end{align}
The decay and inverse decay processes automatically take the resonant part of the $\Delta L = 2$ scatterings into account. Thus to avoid double counting it is a mandatory task to properly subtract the real intermediate states (RIS) contribution where the decaying particle can go on-shell in the s-channel scattering. For a detailed analytical understanding of RIS subtraction one may look into \cite{Buchmuller:2004nz}.  At the same time, it is to note that at a higher temperature the non-resonant parts of $\Delta L = 2$ scatterings become important when the mediating particle (here the scalar $\phi$) is exchanged through u-channel. An in-depth study of such high temperature affect on the $\Delta L = 2$ scatterings mediated by heavy RHNs can be found in \cite{Davidson:2008bu,Giudice:2003jh}.
Now the inverse decay (ID) width $\Gamma_{\rm ID}$ is connected to $\Gamma_{D}$ as:
 \begin{align}
  \Gamma_{\rm ID}(z_i)=\Gamma_D(z_i)\frac{N_{\phi_i}^{\rm eq}(z_i)}{N_l^{\rm eq}},
 \end{align}
where $N_{\phi_i}^{\rm eq}~=~\frac{3}{8}z_i^2\mathcal{K}_2(z_i)$ and $N_l^{\rm eq}=\frac{3}{4}$. Then it follows that the relevant wash out term in the present scenario will take the following form:
\begin{align}
W_i \approx W_i^{\text{ID}} &=\frac{1}{2}\frac{\Gamma_{\rm ID}(z_i)}{Hz},\\
&=\frac{1}{4}K_i {x_{1i}^2}~\mathcal{K}_1 (z_i)z^3,
\end{align}
for standard Universe. We would like to mention once again that in the BEQs of Eq.(\ref{eq:BeL}) $N_{\phi_{i}}$ and $N_{B-L}$ denote the respective abundances with respect to photon number density in highly relativistic thermal equilibrium.

The influence of non-standard cosmology as briefed in Section \ref{fastexpanding}, is observed in the form of a new set of modified 
BEQs where the Hubble rate of expansion obeys the form as shown in Eq.(\ref{modifiedH}). Hence in the alternative cosmological scenario with $n>0$ the Hubble parameter in the present section will be modified according to Eq.(\ref{modifiedH}) wherever applicable.
For example with the new Hubble expansion rate, the decay term looks like,  
\begin{equation}
D_i = \frac{\Gamma_{D, i}}{H z} = K_i z^{n/2+1}x_{1i}^{n/4+1} \frac{\mathcal{K}_1(z_i)}{\mathcal{K}_2(z_i)}.
\end{equation}
Similarly, the washout parameter $K_i$ and $W_{\rm ID} $ will be modified to
\begin{align}
 &K_i=\frac{\tilde{\Gamma}_{D,i}}{H_R(T=M_{\phi_i})}\left(\frac{T_r}{M_{\phi_i}}\right)^{n/2},\\
 &W_i=\frac{1}{4}K_i x_{1i}^{n/4+2}~\mathcal{K}_1 (z_i) z^{n/2+3}
\end{align}
With all these inputs, the final baryon asymmetry of the Universe can be obtained by using,
\begin{equation}
\eta_B = a_{\text{sph}}\frac{N_{\text{B-L}}}{N_\gamma^{\text{rec}}} = 0.0126 ~N_{\text{B-L}}^f,
\end{equation}
where $a_{\rm sph}$ indicates standard sphaleron factor and $N_{\text{B-L}}^f$ being the final B-L asymmetry.

 \begin{table}[b]
\begin{center}
\begin{tabular}{|c| c | c | c | c | c |}
  \hline
  BP & $\Lambda_{11}$ (eV)& $\Lambda_{22}$ (eV)& $\Lambda_{33}$ (eV)& ~$~a~$~ & $h_{\alpha i}\times 10^4$ \\
  \hline
  I & $9.94\times 10^7$ & $1.02\times 10^8$  &$1.04\times 10^8$ & 2.9 &$\left(\tiny
\begin{array}{ccc}\
 -10.08-3.17 i & 4.02\, -7.94 i & -0.31-6.58 i\\
 -1.54\, -10.38 i & 8.92\, 0.26 i & 5.71\, -3.1 i \\
 1.05\, -6.88 i & 5.65\, +1.81 i & 4.13-0.83 i \\
\end{array}
\right)$\\
  \hline
  II & $5.54\times 10^7$  & $5.69\times 10^7$ & $5.83\times 10^6$ & 2.7 &$\left(\tiny
\begin{array}{ccc}
 -9.55-3.0 i & 3.97\, -7.5 i & -0.29-6.22 i \\
 -1.46\, -9.84 i & 8.44\, +0.24 i & 5.39\,- 2.91 i \\
 0.96\, -6.53 i & 5.36\, +1.69i & 3.86-0.75 i \\
\end{array}
\right)$ \\
  \hline
\end{tabular}
\caption{Numerical estimation of the two Yukawa coupling matrices which are obtained for the sets of benchmark points (BP) tabulated in Table~\ref{tab:tab2}. Reference scalar masses are considered as $M_{\phi_i} = \{10^7, 10^{7.1},10^{7.2}\}$ GeV.}
\label{tab:tab3}
\end{center}
\end{table}

\section{Results for neutrino mass and leptogenesis}\label{resultsL}

It is clear from the above discussion that the Yukawa couplings and the masses of BSM scalar and fermionic fields enter into both one loop diagrams responisble for neutrino mass and lepton asymmetry calculation respectively. Here we present some numerical estimates of the relevant parameters which offer correct order of neutrino mass and lepton asymmetry in this set up. 

 For numerical computation we choose the lightest active neutrino mass to be 0.001 eV, abiding by the cosmological bound on the sum of neutrino masses as reported by Planck $\left(\sum_i m_{\nu_i}< 0.12 {\rm ~eV}\right)$ \cite{Aghanim:2018eyx,Giusarma:2016phn,Vagnozzi:2017ovm}. We also prefer to choose the maximal value for Dirac CP phase $\delta_{CP}=-\frac{\pi}{2}$ and the best fit central values for rest of the oscillation parameters. Using these values, it is trivial to obtain the Yukawa couplings ($h_{\alpha i}$) with the help of Eq.(\ref{eq:casas}) once the mass scales of the BSM fields are known. In Table \ref{tab:tab3}, we provide the numerical estimate of the Yukawa couplings matrix ($h$) for the two reference points as noted in Table \ref{tab:tab2}, considering scalar masses as $\{10^7,10^{7.1},10^{7.2}\}$ GeV. This estimation is essential for the calculation of baryon asymmetry as well. 
 
\begin{figure}
\hspace{-1cm}
\includegraphics[height=6.5cm,width=8cm]{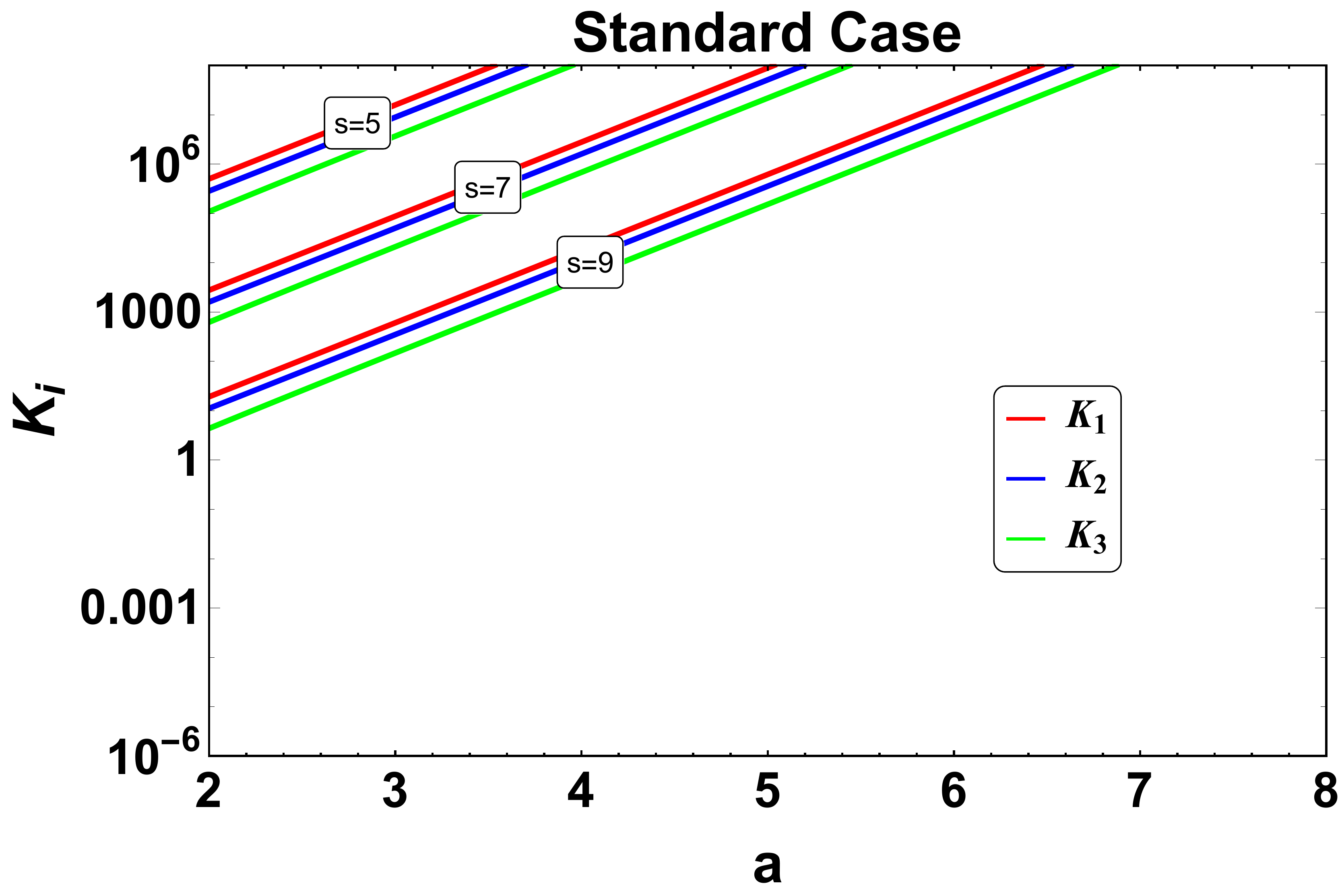}~~
\includegraphics[height=6.5cm,width=8cm]{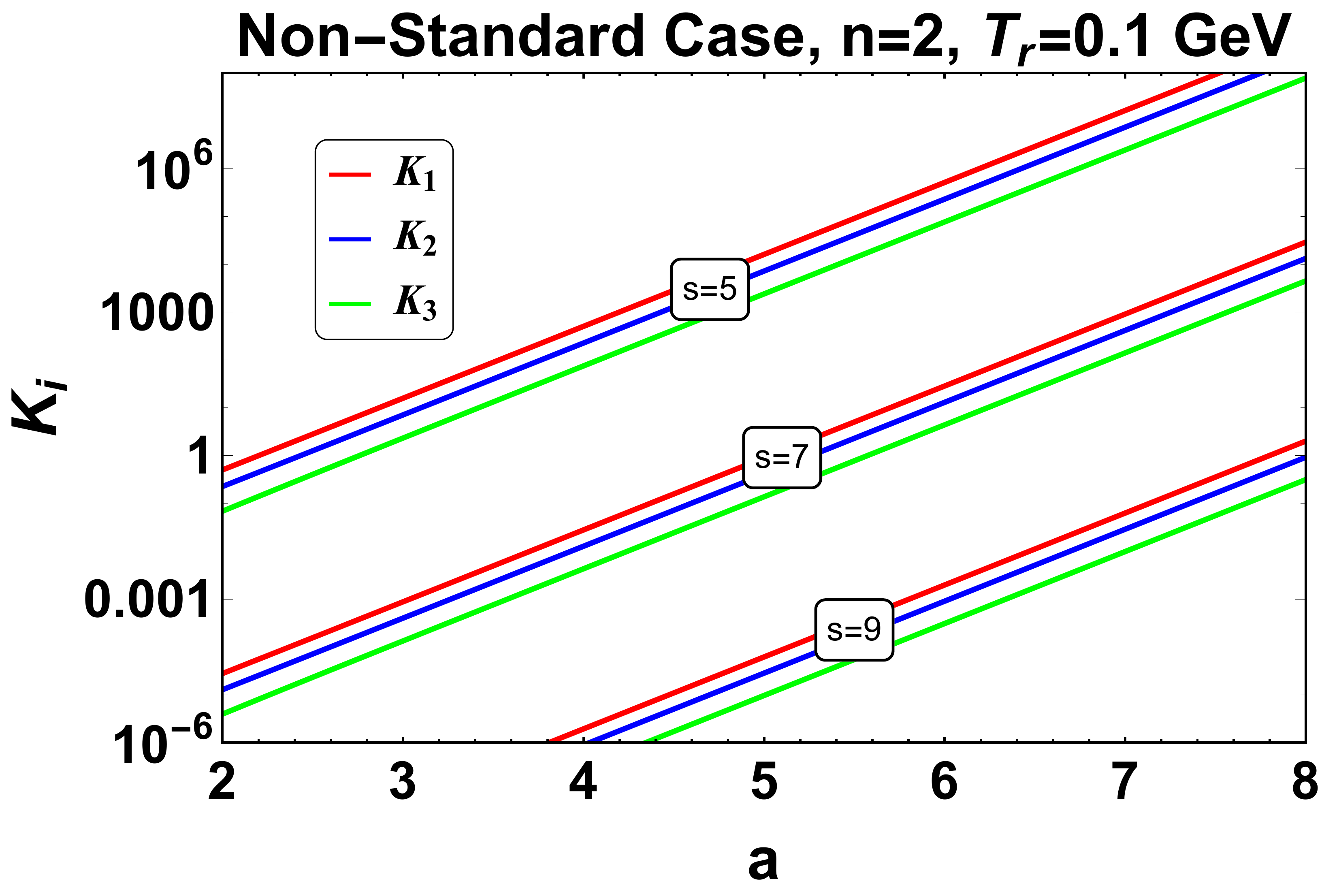}   \\ 
\caption{Washout factors as a function of $a$ for (left) standard and (right) non-standard case. We consider here $M_{\phi_i} = \{10^s, 10^{s+0.1},10^{s+0.2}\}$ GeV with $s=\{7,8,9\}$ for the benchmark point I in Table \ref{tab:tab2}.}
\label{fig:Wash1} 
\end{figure}
\begin{figure}
\centering
\hspace{-5mm}\includegraphics[height=5.5cm,width=7.5cm]{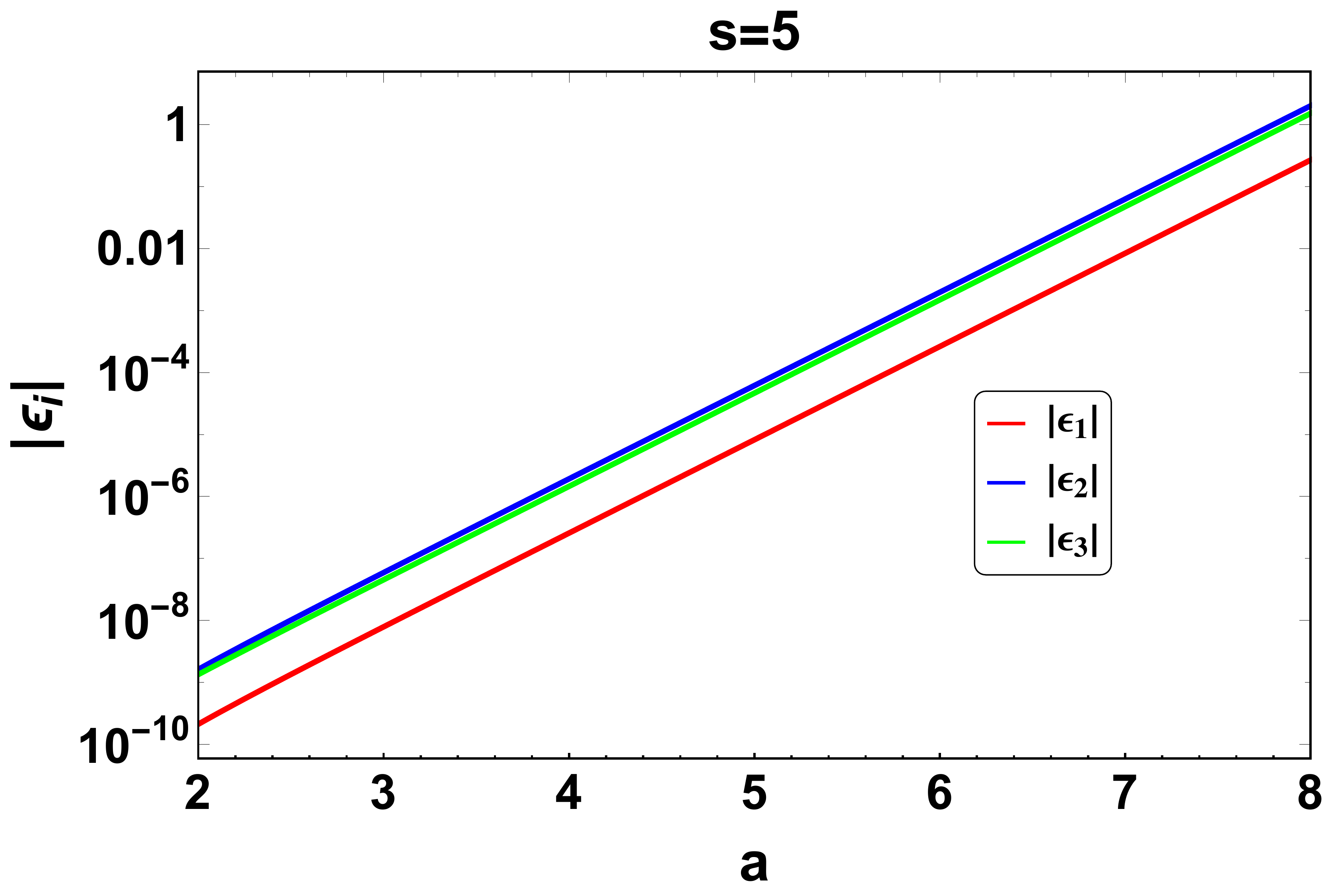}
\includegraphics[height=5.5cm,width=7.5cm]{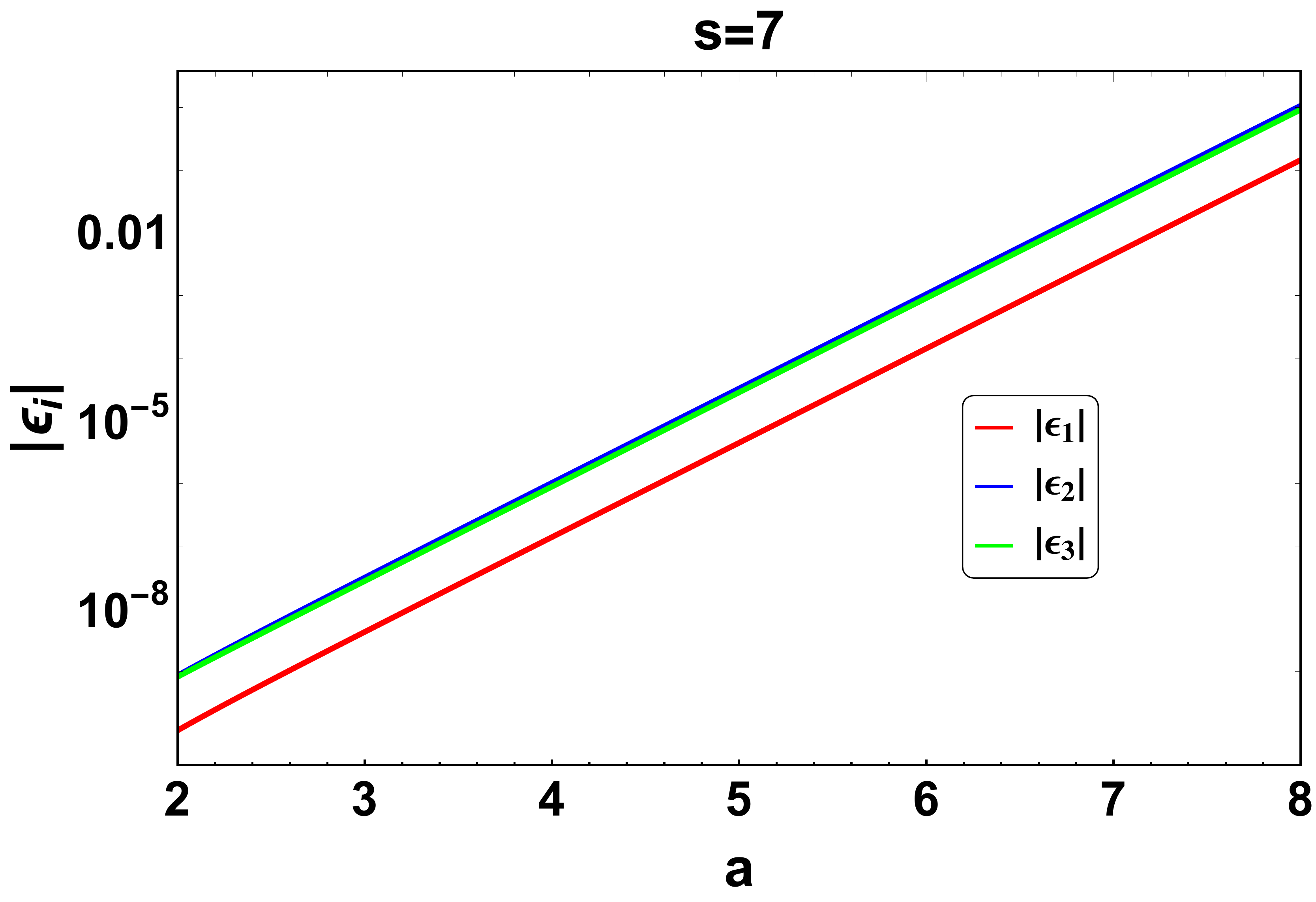}
\caption{Order of lepton asymmetry parameter $\epsilon_i$ as a function of $a$ in Eq.(\ref{eq:casas}) for considering scalar masses $M_{\phi_i} = \{10^5, 10^{5.1},10^{5.2}\}$ GeV (left) and $M_{\phi_i} = \{10^7, 10^{7.1},10^{7.2}\}$ GeV (right) for the benchmark point I in Table \ref{tab:tab2}.}
\label{fig:Wash2} 
\end{figure}

 As emphasized earlier, one of the primary aims of this study is to investigate the dynamical generation of baryon asymmetry considering the presence of non-standard cosmology ($H\neq H_R$) instead of the standard one ($H=H_R$). The Figs. \ref{fig:Wash1}-\ref{fig:Wash2} illustrate the reason behind this preference. In Fig. \ref{fig:Wash1}, we show the variation of the washout factor $K_i$ as a function of the parameter $a$ present in Eq.(\ref{eq:casas}) considering both standard (left) and non-standard (right) cases. In Fig. \ref{fig:Wash2}, we exhibit the variation of $\epsilon_i$ with respect to the parameter $a$. For clarity we have chosen different domains for the scalar mass, considering $M_{\phi_i} :\{10^s,10^{s+0.1},10^{s+0.2}\}$ GeV where $s$ can take the values as $s={5,7,9}$. Using this set of $M_{\phi_i} $ values and the reference point I in Table \ref{tab:tab2} we prepare these figures. These figures give a clear insight on the fact that both the washout factor $K_i$ and $\epsilon_i$ are increasing functions of $a$. Moreover, for lower $M_{\phi_i}$ the wash out becomes stronger ($K_i\gg 1$). The Fig. \ref{fig:Wash2} reveals that the order of the asymmetry parameter remains to be more or less unaltered irrespective of the choice of $M_{\phi}$ scales. This can be understood from Eq.(\ref{eq:ep1}), where the term involving the functional dependence of $M_{\phi_i}$ takes a constant value close to unity for any arbitrary choice of $M_{\phi_i}$.  
 
 In contrast to the standard case, the right  panel of Fig. \ref{fig:Wash1} shows that the order of $K_i$'s can be substantially suppressed in case the Universe expands faster where we have chosen $T_r$ and $n$ to be $0.1$ GeV and $2$ respectively. Although in the standard case it may be possible to generate the correct order of baryon asymmetry with superheavy scalar fields ($M_{\Phi_i}\gg 10^9$ GeV), we prefer the non-standard option since it opens up the possibility of relaxing the lower bound on $M_\phi$'s to meet the weak washout criteria ($K_i< 1$).   

\begin{figure}
\hspace{-5mm}
\includegraphics[height=6cm,width=7.5cm]{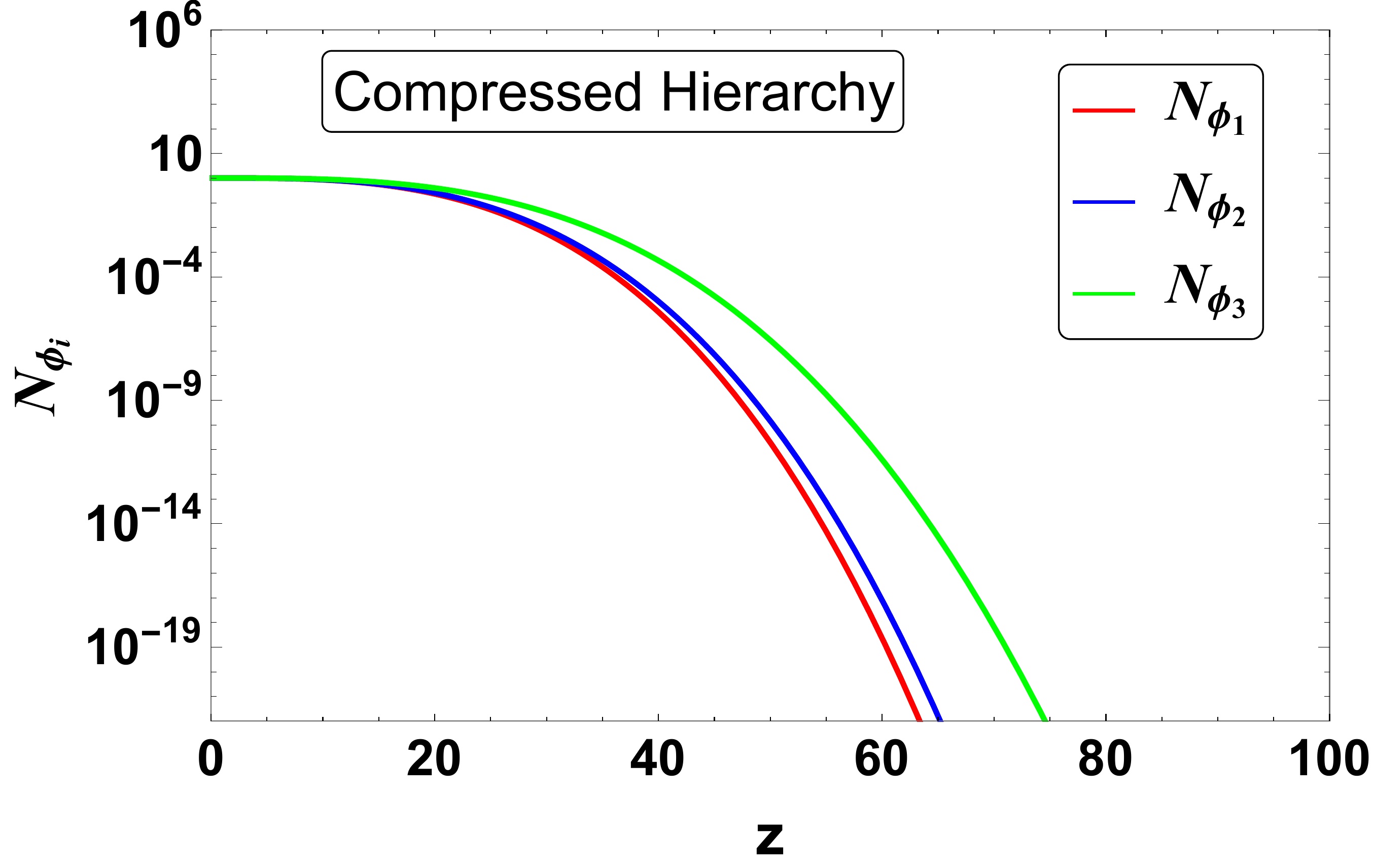}
\includegraphics[height=6cm,width=7.5cm]{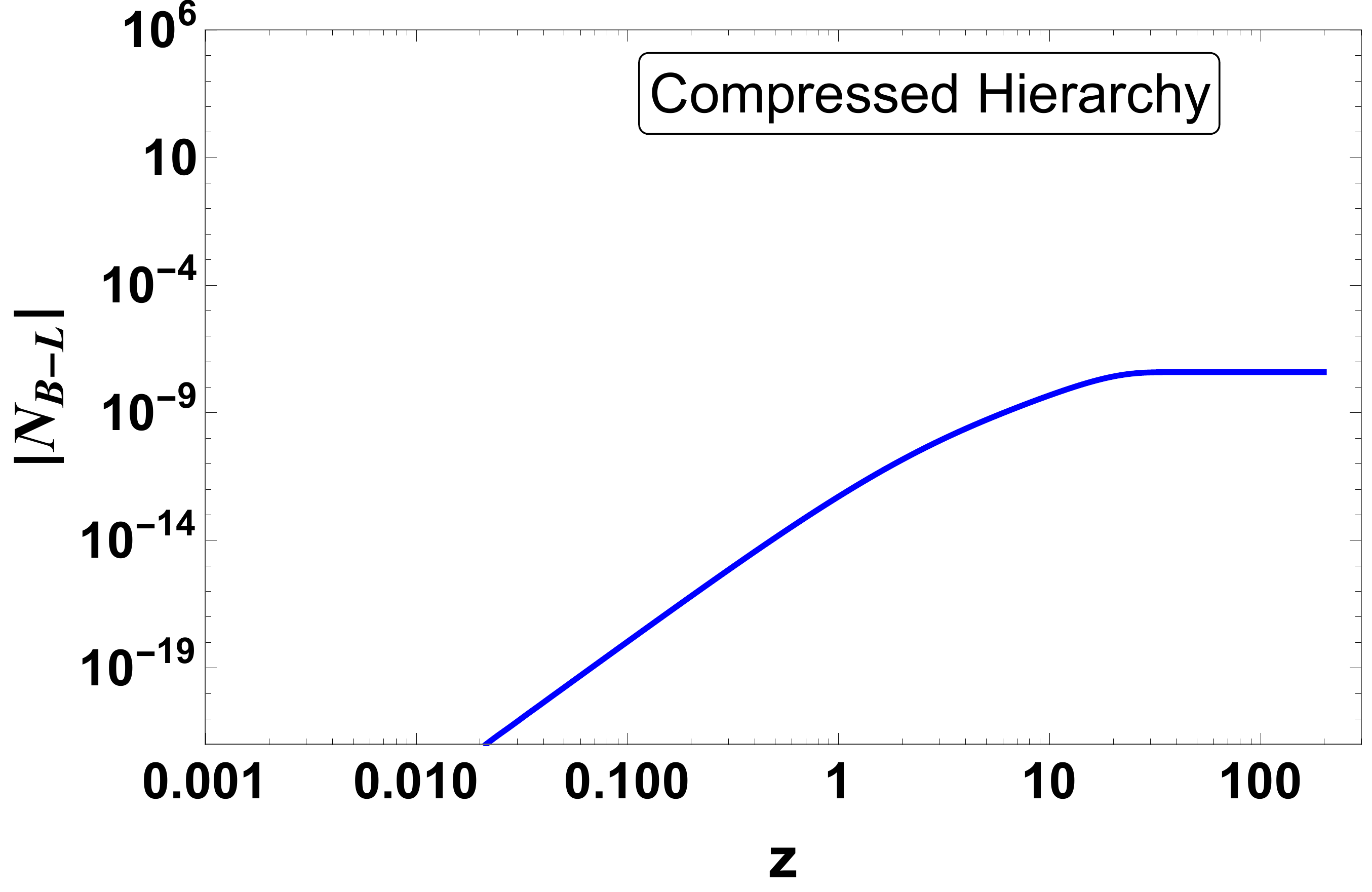}
\caption{Evolution of $N_{\phi_i}$ (left) and $N_{\rm B-L}$ (right) as a function of temperature $T$ considering compressed mass hierarchy among the scalars with $\{M_{\phi_i}\rightarrow 10^{7},10^{7.1}, 10^{7.2}\}$ GeV and $T_r=0.1$ GeV for the benchmark point I in Table \ref{tab:tab2}.} 
\label{fig:No} 
\end{figure}

\begin{figure}
\hspace{-5mm}
\includegraphics[height=6cm,width=7.5cm]{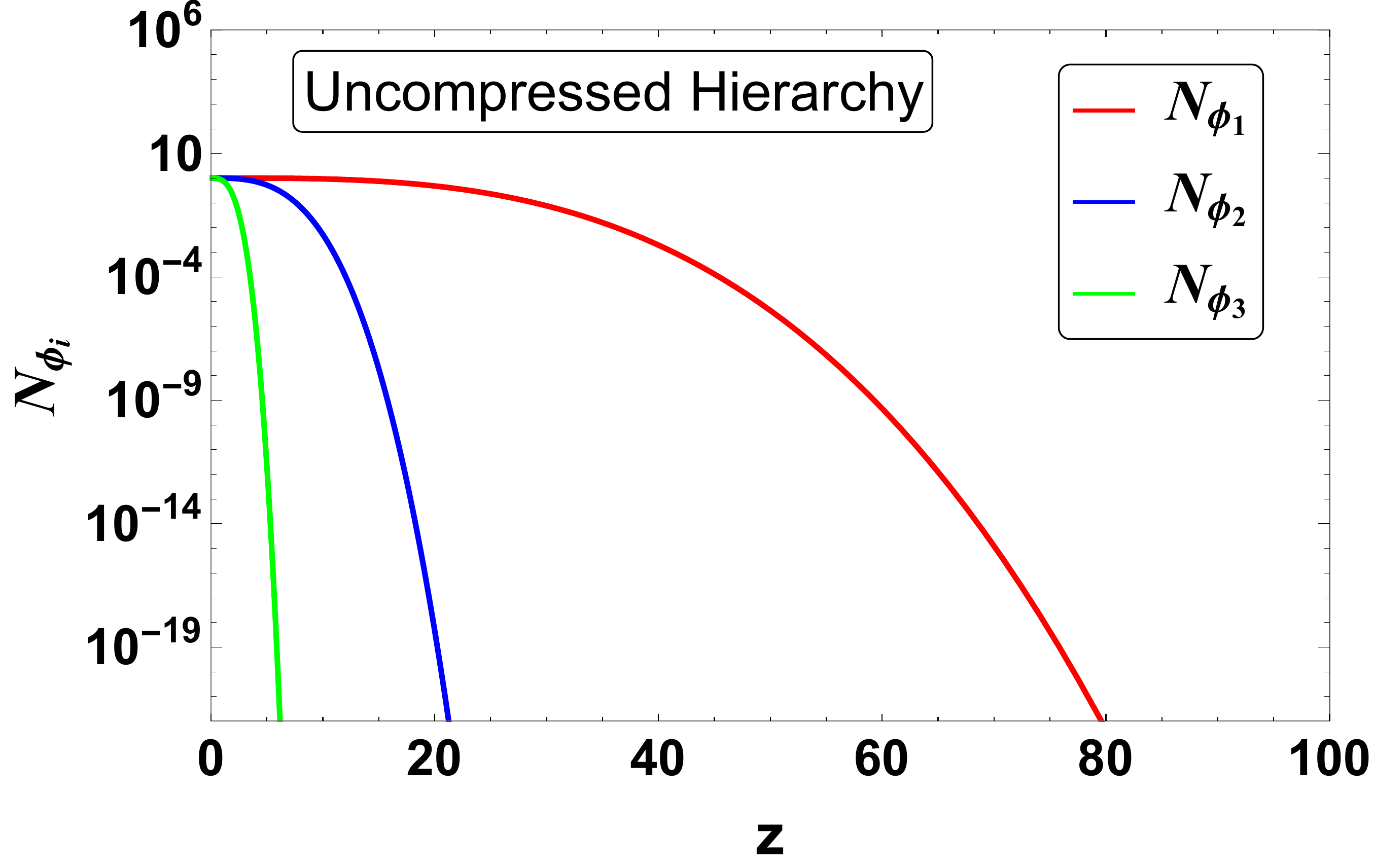}
\includegraphics[height=6cm,width=7.5cm]{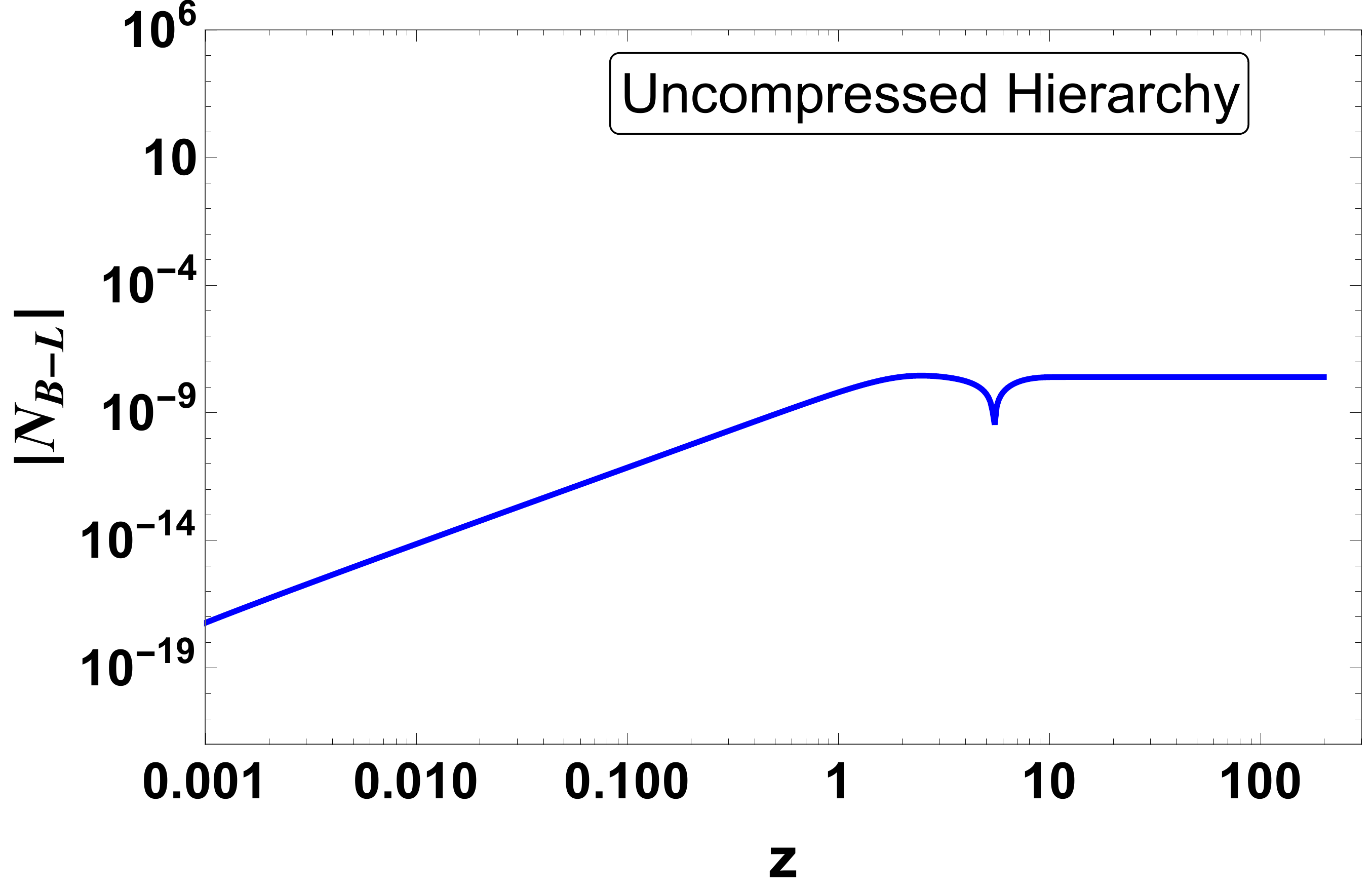}
\caption{Evolution of $N_{\phi_i}$ (left)  and $N_{\rm B-L}$ (right) as a function of temperature $T$ considering uncompressed mass hierarchy among the scalars with $\{M_{\phi_i}\rightarrow 10^{7},10^{9}, 10^{11}\}$ GeV and $T_r=0.1$ GeV for the benchmark points I in Table \ref{tab:tab2}.}
\label{fig:No1} 
\end{figure}

We numerically solve the BEQs of Eq.(\ref{eq:BeL}) with the initial conditions that the scalars are in thermal equilibrium at $T>M_{\phi_i}$ and also assume that the initial B-L asymmetry $N_{B-L}^{\rm ini}=0$. We have performed this analysis by assuming the lightest scalar $M_{\phi_1}\sim \mathcal{O}(10^7)$ GeV and which is enforced to obey two kinds of hierarchies with the other two heavier scalars. First we consider a compressed pattern of mass hierarchy among the scalars and in the later part we speculate on the case with a relatively larger mass hierarchy. This two hierarchy patterns lead to distinct evolutionary dynamics of the scalars as understood from Figs. \ref{fig:No}-\ref{fig:No1}.

In Fig. \ref{fig:No}, we show the evolution of $N_{\phi_{1,2,3}}$ (left) and $N_{B-L}$ (right) by considering the compressed mass pattern with $n=2, ~M_{\phi_i}=\{10^{7},10^{7.1},10^{7.2}\}$ GeV. As it is seen that, number density of the scalars drops from their equilibrium abundances and $N_{\rm B-L}$ rises with decreasing temperature and finally $N_{\rm B-L}$ gets saturated at some finite value. In Table \ref{tab:tab4}, we list the required values of the parameter $a$ to attain the observed amount of $\eta_B$ for the reference points of Table \ref{tab:tab2} considering $n=2$ and $T_r=0.1$ GeV. We also include the order of the lepton asymmetry parameter and the $\eta_B$ values for $n=1$. It is clearly understood that a smaller value of $n$, reduces the amount of $\eta_B$ for a fixed $T_r$ and $a$. 

Next we consider a representative uncompressed mass hierarchies among the scalars (not shown in the Tables) and fix $M_{\phi_i}=\{10^7, 10^9, 10^{11}\}$ GeV. In Fig. \ref{fig:No1}, we show the evolution of $N_{\phi_{1,2,3}}$ and $N_{B-L}$ as a function of temperature $T$. Since $M_{\phi_{2,3}}$ are quite heavier as compared to $M_{\phi_1}$, their number densities fall sharply at a very early stage of evolution. Hence, in the evolution, first $N_{B-L}$ gets created from $\phi_3$ decay. Then when $\phi_2$ starts decaying, $N_{B-L}$ changes its sign which is observed in form of a kink in right of Fig. \ref{fig:No1} is observed. Finally the decay of the lighter scalar $\phi_1$ helps in keeping the remnant asymmetry upto the expected amount successfully. Similar to the earlier case, in Table \ref{tab:tab5}, we tabulate the findings: the value of $a$, order of $\epsilon_{1,2,3}$ and $\eta_B(n=1)$ to attain the correct order of $\eta_B$. 

\begin{table}[]
\begin{center}
\begin{tabular}{| c | c | c | c | c | c | c |}
  \hline
BP & ~~$a$~~ & $|\epsilon_1|$ & $|\epsilon_2|$ & $|\epsilon_3|$  & $\eta_B(n=1)$ & $\eta_B(n=2)$  \\
  \hline
  I & 2.9 & $3.01\times10^{-9}$ & $2.21\times 10^{-8}$   & $1.95\times10^{-8}$  & $5.07\times 10^{-13}$& $3.71 \times 10^{-10}$   \\ 
  \hline
  II & 2.7 & $2.68\times 10^{-9}$  & $1.98\times 10^{-8}$  &  $1.7\times 10^{-8}$   & $5.69\times10^{-13}$ & $3.28 \times10^{-10}$\\
  \hline
\end{tabular}
\caption{Estimating baryon asymmetry considering compressed mass hierarchy with $M_{\phi_i}=\{10^7,10^{7.1},10^{7.2}\}$ GeV for the two benchmark points in Table \ref{tab:tab2}.}.
\label{tab:tab4}
\end{center} 
\end{table}
\begin{table}
\begin{center}
\begin{tabular}{| c | c | c | c | c | c | c |}
  \hline
  BP & ~~$a$~~ & $|\epsilon_1|$ & $|\epsilon_2|$ & $|\epsilon_3|$ &   $\eta_B(n=1)$ & $\eta_B(n=2)$  \\
  \hline
  I & 2.75 & $8.99\times 10^{-13}$ & $7.03\times 10^{-8}$   & $ 4.05\times10^{-8}$    & $1.30\times10^{-13}$& $2.86 \times 10^{-10}$   \\
  \hline
  II & 2.75 & $1.52\times 10^{-12}$  & $1.26\times 10^{-7}$  &  $8.16\times 10^{-8}$   & $2.10\times10^{-16}$ & $4.30\times10^{-10}$\\
  \hline
\end{tabular}
\caption{Estimation of baryon asymmetry considering uncompressed mass hierarchy with $M_{\phi_i}=\{10^7,10^9,10^{11}\}$ GeV for the two benchmark points in Table \ref{tab:tab2}.}.
\label{tab:tab5}
\end{center} 
\end{table}

The present analysis appears to be suitable for any mass window for the scalars provided the validity of the analytical  expressions for $\epsilon_{1,2,3}$ in Eq.(\ref{eq:ep1}) holds. It is to note here that, as of now we have explored this scenario only for unflavored regime of leptogenesis, but it would be intriguing to examine this framework including flavor effects where the charged lepton Yukawa interactions are fast enough. In analogy with the scenario where lepton asymmetry originates from the decay of a heavy RHN, a different magnitude of $a$ would be required to describe the evolution of such processes, consistent with the observations. Since, we are already in the weak wash out regime, apparently it can be claimed that the contribution from the individual flavor asymmetries would be minimum \cite{Davidson:2008bu}. 

\section{Summary and Conclusion}\label{conclusion}
 We have constructed an attractive framework deciphering baryogenesis from leptogenesis along with a pseudo-Dirac dark matter candidate and neutrino mass in a scalar extended singlet doublet scenario. Successful accomplishment of all the three entities at the same time is conspired by a mere Majorana mass term for the singlet fermion present in the Lagrangian. We have considered both standard and non-standard cosmology and furnished a comparative analysis between the two. Since the thermal history of the Universe is largely unknown prior to the big bang nucleosynthesis, we conceive the idea of fast-expanding Universe and analyze the singlet doublet DM phenomenology in detail. In one of our earlier works, we have investigated pseudo-Dirac singlet doublet DM phenomenology in view of spin independent DM SI direct search experiments. Here we extend that idea and find that the impact of this rapid expansion of the Universe turns significant especially the relevant parameter space to be consistent with the direct detection bound receives huge deviation compared to the standard one. First, we estimate the interaction strength for the singlet doublet dark matter with the visible sector for two specific DM masses ($\lesssim 1$ TeV) considering the various kinds of the fast expansion of the Universe (with different temperature dependences) which turns out to be higher than in the usual scenario. This looks consistent with the earlier model independent works in this direction.
In the later part, we discuss the radiative generation of neutrino mass which require an extension of the minimal framework with additional singlet scalars. We further calculate the baryon asymmetry of the Universe from the decay of these dark scalars by using the Yukawa couplings which get constrained from the neutrino oscillation data. The proposed mechanism of lepton asymmetry generation is slightly different from the ones available in the existing literature where the decay of heavy right-handed neutrino generates the asymmetry in the lepton sector. We conclude with an important notion that the non-standard Universe is perhaps preferred over the standard one in the present scenario to yield the observed amount of baryon asymmetry in the Universe.


\section*{Acknowledgements}\label{ack}
This work is supported by Physical Research Laboratory (PRL), Department of Space, Government of India. Computations were performed using the HPC resources  (Vikram-100 HPC) and TDP project at PRL. Authors gratefully acknowledge WHEPP'19 where parts of this work were initiated. AM likes to appreciate  M. Ratz for helpful conversation. Authors also thank KM Patel and S Seth for useful  discussion. AKS is partially supported by Science and Engineering Research Board, Government of India, under NPDF grant PDF/2020/000797.

\appendix
\section{{ Realization of Pseudo-Dirac fermion}}\label{pseudo}
{ 
Here, we present a brief understanding on the construction of a pseudo-Dirac fermion. The notations to be used are adopted from ref.~\cite{Dreiner:2008tw}. In general a Dirac fermion ($X$) can be expressed in terms of two Weyl fermions.
\begin{align}
X=\begin{pmatrix}
\eta\\
\xi^\dagger
\end{pmatrix} {\rm ~~~~with ~~~} P_L X=\begin{pmatrix}
\eta\\
0
\end{pmatrix} {\rm ~~~~and ~~~~} P_R X= \begin{pmatrix}
0\\
\xi^\dagger
\end{pmatrix}
\end{align}
The charge conjugation of $X$ is given by
\begin{align}
X^c=\begin{pmatrix}
\xi\\
\eta^\dagger.
\end{pmatrix}
\end{align}
Suppose we have a Lagrangian where both Dirac and Majorana mass terms for $X$ field are present.
\begin{align}
\mathcal{L}=m_D\bar{X}X+\frac{1}{2}m_M^L\overline{X^C}P_LX+h.c.+\frac{1}{2}m_M^R\overline{X^C}P_RX+h.c..
\end{align}
For the moment we consider $m_M^{L,R}=m_M$. Now, one can expand $\mathcal{L}$ in terms of the Weyl components and can rewrite the above Lagrangian as,
\begin{align}
\mathcal{L}=m_D(\eta\xi+\eta^\dagger\xi^\dagger)+\frac{1}{2}m_M(\eta\eta+h.c.)+ \frac{1}{2}m_M(\xi\xi+h.c.)
\end{align}
With this, the mass matrix in the $(\eta,\xi)$ basis turns out to be
\begin{align}
\mathcal{M}=\begin{pmatrix}
m_M & m_D\\
m_D & m_M
\end{pmatrix}.
\end{align}
After diagonalising $\mathcal{M}$, we obtain the two mass eigenstates as,
\begin{align}
&y_1=\frac{i}{\sqrt{2}}(\xi-\eta)\\
&y_2=\frac{1}{\sqrt{2}}(\xi+\eta).
\end{align}
We notice that the eigenstates $y_1$ and $y_2$ are two component in nature. However, we can always form a four component spinor by
defining 
\begin{align}
Y_{i}=\begin{pmatrix}
y_{i}\\
y_{i}^\dagger
\end{pmatrix}.
\end{align}
With the above definition it is convenient to write the following relations,
\begin{align}
&Y_1=\frac{i}{\sqrt{2}}(X^c-X)\\
&Y_2=\frac{1}{\sqrt{2}}(X^c+X).
\end{align}
It is important to mention that, in the limit $m_M^L\neq m_M^R$, the above four component eigenstates will receive some correction which can be expressed as
\begin{align}
&Y_1=\frac{i}{\sqrt{2}}(X^c-X)+\mathcal{O}(\delta)\\
&Y_2=\frac{1}{\sqrt{2}}(X^c+X)+\mathcal{O}(\delta),
\end{align}
where, $\delta=|m_M^L-m_M^R|$. Now, the states $Y_1$ and $Y_2$ are non degenerate and since they have a pseudo-Dirac origin, we call them pseudo-Dirac states in the limit $m_M^{L,R}\ll m_D$.} 

\section{Analytical formulation of the lepton asymmetry parameter}\label{appen1}
In this section we present a brief analytical estimate of the lepton asymmetry from the lepton number violating dark sector scalar singlet decay. The asymmetry parameter generally gets non-zero contributions from the interference of the tree level and two 1-loop level diagrams as shown in Fig. \ref{feyn}.  However in the present set up with a vanishing lepton mass limit, the sole contribution to the lepton asymmetry is sourced by the interference of the tree level and vertex diagram only.  The invariant  amplitude square for the tree level decay of the BSM scalar ($\phi$) to SM lepton ($l$) and the vector like fermion ($\Psi$) can be expressed as,
\begin{equation}
|\mathcal{M}|^2_{~\phi_i \rightarrow~ \bar l_{L_{\alpha}} + \Psi} =  \sum_{\alpha}\left(h_{\alpha i}h^*_{\alpha i}\right)M_{\phi_i}^2,
\end{equation}
Where $i,~ j$ are the indices specific to the BSM scalar which run as $(1,2,3)$ and the $\alpha = e,~\mu,~\tau$ refers to SM lepton indices respectively. Considering the limit $ M_\phi \gg m_\Psi, m_l$, the corresponding decay width of $i$'th scalar at tree level can be expressed as:
\begin{align}
\Gamma_{\phi_i \rightarrow~ \bar l_{L_{\alpha}} + \Psi} &=\frac{|\mathcal{M}|^2}{16 \pi}\frac{1}{M_{\phi_i}} \nonumber \\
&=\frac{\left(h^{\dagger}h\right)_{ii}M_{\phi_i}}{16\pi}.\label{eq:treedecay}
\end{align}

\begin{figure}[t]
\centering
\includegraphics[height=4cm,width=6cm]{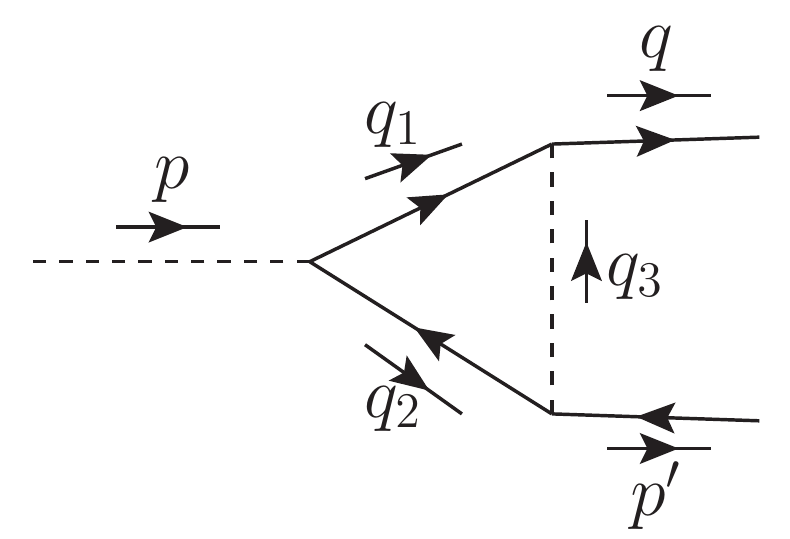}
\caption{Particle directions and momenta of the vertex diagram as shown in Fig. \ref{feyn}.}
\label{appendix1}
\end{figure}

Next we proceed to calculate the contribution caused by the interference between the tree level and vertex diagrams to $\epsilon$. The Feynman amplitude square of such kind of interference process (see Fig. \ref{appendix1}) is given by (in the vanishing lepton and DM mass limits):
\begin{align}
I_{\rm vertex}^\prime= 2i A_h\int \frac{d^4q_1}{(2\pi)^4}\frac{\overline{u_\Psi}~P_R~\cancel{q_1}~P_L\cancel{q_2}~P_R ~v_l~\overline{v_l}~P_L ~u_\Psi}{(q_1^2+i\varepsilon)(q_2^2+i\varepsilon)(q_3^2-M_{\phi_j}^2+i\varepsilon)},\label{eq:intL1}
\end{align}
where $A_h=h_{\beta j}h^*_{\beta i}h_{\alpha j} h^*_{\alpha i}$.
Afterwards, we use the standard trace properties of the Gamma matrices and also consider the imaginary part of the $I_{\rm vertex}^\prime /A_h$ in Eq.(\ref{eq:intL1}) (since it solely matters for lepton asymmetry as we will see in a while) to write:
\begin{align}
I_{\rm vertex}^\prime=2 i A_h \int \frac{d^4q_1}{(2\pi)^4}\frac{(q_1.q_2)(p^\prime.q)-(q_1.p^\prime)(q_2.q)+(q_1.q)(q_2.p^\prime)}{(q_1^2+i\varepsilon)(q_2^2+i\varepsilon)(q_3^2-M_{\phi_j}^2+i\varepsilon)},
\end{align} 
 We work in rest frame of the incoming particle $\phi_i$. Applying principle of momentum conservation at each vertices we obtain
\begin{align}
q_1=\{E_1,{\bf q_1}\},~~ q_2=\{M_{\phi_i}-E_1,-{\bf q_1}\}~ {\rm ~with~} p=\{M_{\phi_i},\vec{0}\},\\
{\rm also,~~} p^\prime=\left\{\frac{M_{\phi_i}}{2},-{\bf q}\right\}~~,q=\left\{\frac{M_{\phi_i}}{2},{\bf q}\right\}.
\end{align}
Next, we implement the famous Cutkosky rule to evaluate the integral Im$(I_{\rm vertex})$ and write
\begin{align}
{\rm Disc}~[I_{\rm vertex}^\prime]=2 i A_h \int \frac{d^4 q_1}{(2\pi)^4} \frac{(-2\pi i)^2~\delta[q_1^2]~\delta[(p-q_1)^2]~\Theta(E_1)~\Theta(M_{\phi_i}-E_1)}{\left[(q-q_1)^2-M_{\phi_j}^2\right]}\label{eq:mla1}
\end{align}
Upon further simplifications and performing the integral Eq.(\ref{eq:mla1}) we reach at
\begin{align}
{\rm Disc}~[I_{\rm vertex}^\prime]=\frac{i A_h  M_{\phi_i}^2}{8\pi}\left[1+x_{ij} ~\text{~log}\left(\frac{x_{ij}}{1+x_{ij}}\right)\right],
\end{align}
where $x_{ij}=\frac{M_j^2}{M_i^2}$. Now, one can use the conversion: ${\rm Im}\left(I^\prime\right)=-\frac{1}{2i}~{\rm Disc}~[I^\prime]$ to attain
\begin{align}
{\rm Im}\left(I^\prime\right)=-\frac{M_{\phi_i}^2}{16\pi}\left[1+x_{ij} ~\text{~log}\left(\frac{x_{ij}}{1+x_{ij}}\right)\right].
\end{align}
where we define $I_{\rm vertex}^\prime= A_h I^\prime$.
The general formula for vertex contribution to the lepton asymmetry parameter is,
\begin{equation}
\epsilon_{\text{vertex}} = -\frac{4}{\Gamma_{\text{tot}}}\sum_{i\neq j}\sum_\alpha \text{Im}(A_h)\text{~Im}[I^\prime V_\phi],
\end{equation}
where $V_\phi$ is the phase space factor for a two body decay process (under discussion) having magnitude $\frac{1}{8\pi M_{\phi_i}}$. The total decay width is the sum of forward and inverse decay widths {\it i.e.} $\Gamma_{\rm tot}=\Gamma_{\phi_i}+\bar{\Gamma}_{\phi_i}$ as in Eq.(\ref{eq:treedecay}). One can further write $\text{Im}(I^\prime V_\phi) =\text{Im}(I^\prime) V_\phi$ since $V_\phi$ is real.

With all the expressions earlier highlighted, finally we note down the explicit form of $\epsilon_{\rm vertex}$ in terms of the model parameters,
\begin{equation}
 \epsilon_{\text{vertex}}^i = \frac{1}{4\pi}\sum_{j\neq i}\frac{\text{Im} \left[(h^\dagger h)_{ij}h_{\alpha j}h_{\alpha i}^*\right]}{(h^\dagger h)_{ii}}\left[1+x_{ij} \text{~log}\left(\frac{x_{ij}}{x_{ij}+1}\right)\right]
\end{equation}

In a similar fashion, one can formulate the contribution to the lepton asymmetry originating from the self energy diagram. Since at vanishing lepton mass limit due to properties of Gamma matrices, the interference amplitude of self energy and the tree level diagrams vanishes at the amplitude level we skip the details here.

\bibliographystyle{JHEP}
\bibliography{refs1}

\end{document}